%% file: mct.tex
\documentclass[final]{JHEP3} 

\JHEPspecialurl{http://jhep.sissa.it/JOURNAL/JHEP3.tar.gz}

\usepackage{epsfig,multicol,bbm,amsmath,amssymb,ifpdf}

\newcommand\fverb{\setbox\fverbbox=\hbox\bgroup\verb}
\newcommand\fverbdo{\egroup\medskip\noindent%
			\fbox{\unhbox\fverbbox}\ }
\newcommand\fverbit{\egroup\item[\fbox{\unhbox\fverbbox}]}
\newbox\fverbbox

\def\chioi{\tilde{\chi}^0_1}

\def\chipm{\tilde{\chi}^{\pm}_1}

\def\sqr{\tilde{q}_R}

\def\ra{\rightarrow}

\def\slep{\tilde{\ell}^{\pm}}
\def\snu{\tilde{\nu}}

\def\mct{M_{CT}}
\def\mc{M_{C}}
\def\mtrans{M_T}
\def\mtransmax{M_T^{\rm max}}
\def\mctmax{M_{CT}^{\rm max}}

\def\mctmin{M_{CT}^{\rm min}}
\def\mcto{M_{CT{\rm(CoM)}}}
\def\mctl{M_{CT{\rm(lab)}}}
\def\mctlc{M_{CT{\rm(corr)}}}
\def\mmax{m_{\rm max}}

\def\eog{E_{\delta\delta}^{\rm est}}
\def\ehat{\widehat{E}}
\def\mcy{M_{Cy}}
\def\axl{A_{x({\rm lab})}}
\def\ecm{E_{cm}}
\def\axecm{A'_{x({\rm lo})}}
\def\axmthard{A'_{x({\rm hi})}}
\def\axp{A'_x}

\def\mchioi{m(\chioi)}
\def\mchipm{m(\chipm)}
\def\msl{m(\slep)}
\def\msnu{m(\snu)}

\def\mchioisqr{m^2(\chioi)}
\def\mchipmsqr{m^2(\chipm)}

\def\max{{\rm max}}

\def\MT2{M_{T2}}

\def\m0{${0}$}

\def\tg{{\tilde g}}
\def\tq{{\tilde q}}

\def\tchi{{\tilde\chi}}
\def\tl{{\tilde\ell}}

\def\lsp{{\tilde\chi_1^0}}

\def\mw{m(W)}
\def\mt{m(t)}
\def\mnu{m(\nu)}
\def\mwsqr{m^2(W)}
\def\mtsqr{m^2(t)}
\def\mnusqr{m^2(\nu)}

\title{Supersymmetric particle mass measurement with the
boost-corrected contransverse mass}

\author{Giacomo
	Polesello,\\ INFN, Sezione di Pavia, Via Bassi 6, 27100 Pavia,
	Italy \\ E-mail: \email{giacomo.polesello@cern.ch}}
	\author{Daniel R. Tovey,\\ Department of Physics and
	Astronomy, \\ University of Sheffield, Hounsfield Road,
	Sheffield S3 7RH, UK\\ E-mail:\email{ daniel.tovey@cern.ch}}

\received{} 		
\accepted{}		

\preprint{}

\abstract{A modification to the contransverse mass ($\mct$) technique
for measuring the masses of pair-produced semi-invisibly decaying
heavy particles is proposed in which $\mct$ is corrected for non-zero
boosts of the centre-of-momentum (CoM) frame of the heavy states in
the laboratory transverse plane. Lack of knowledge of the mass of the
CoM frame prevents exact correction for this boost, however it is
shown that a conservative correction can nevertheless be derived which
always generates an $\mct$ value which is less than or equal to the
true value of $\mct$ in the CoM frame. The new technique is
demonstrated with case studies of mass measurement with fully leptonic
$t\bar{t}$ events and with SUSY events possessing a similar final
state. }


\begin{document} 


\input{intro.tex}

\input{theory.tex}

\input{indmass.tex}

\input{twostep.tex}

\input{squark.tex}
\input{conclusions.tex}

\section*{Acknowledgements}
The authors wish to thank Mihoko Nojiri and Alan Barr for helpful
comments. DRT wishes to acknowledge STFC and the Leverhulme Trust for
support.

{\bf Note added to version 2:} Since version 1 of this paper was
released onto arXiv a paper \cite{Matchev:2009ad} has been released
which also derives Eqn.~(\ref{eqn12d}). The boost-dependence study
described in Section~\ref{subsec3.2}, which appeared in version
2 of this paper, was carried out without knowledge of
Ref.~\cite{Matchev:2009ad}, however we are happy to acknowledge that
Eqn.~(\ref{eqn12d}) appeared in Ref.~\cite{Matchev:2009ad} first.

\appendix
\input{appendix.tex}

\end{document}

%% file: intro.tex
\section{Introduction}\label{sec1}

Techniques for measuring the masses of pair-produced particles
decaying semi-invisibly through short decay chains at hadron colliders
have attracted considerable interest. The principle motivation for the
development of such techniques is the measurement of the masses of
supersymmetric particles (`sparticles') at the Large Hadron Collider
\cite{Lester:1999tx,Barr:2003rg,Weiglein:2004hn,Cho:2007qv,Gripaios:2007is,Barr:2007hy,Cho:2007dh,Nojiri:2008hy,Tovey:2008ui,Barr:2008ba,Nojiri:2008vq,Cho:2008tj,Cheng:2008hk,Burns:2008va,Barr:2008hv,Matchev:2009fh},
however they may be applied more widely to measure the mass of the top
quark at the Tevatron \cite{cdfmt2} or LHC \cite{Cho:2008cu}, or to
identify fully leptonic $WW$ events \cite{Barr:2005dz}.

Recently \cite{Tovey:2008ui} a straightforward new variable, the
`contransverse mass' ($\mct$), was proposed which enables the
measurement of a simple analytical combination of the masses of the
pair-produced heavy states $\delta_i$ ($i=1,2$) and their invisible
decay products $\alpha_i$. The contransverse mass is defined by
\begin{eqnarray}
\label{eqn1}
M_{CT}^2(v_1,v_2) & \equiv & [E_T(v_1)+E_T(v_2)]^2 - [{\bf p_T}(v_1)-{\bf p_T}(v_2)]^2\cr
& = & m^2(v_1) + m^2(v_2) + 2[E_T(v_1)E_T(v_2)+{\bf p_T}(v_1) \cdot {\bf p_T}(v_2)],
\end{eqnarray}
where $v_i$ are the visible products of each decay chain, ${\bf
p_T}(v_i)$ is the tranverse momentum vector of $v_i$ and
\begin{equation}
\label{eqn2}
E_T(v_i) \equiv \sqrt{p_T^2(v_i)+m^2(v_i)}.
\end{equation}
It can be shown \cite{Tovey:2008ui} that $\mct$ is in general bounded
from above by a quantity dependent upon the masses $m(\delta)$ and
$m(\alpha)$. If $m(v_1)=m(v_2)\equiv m(v)$ then the
distribution of event $\mct$ values possesses an end-point at:
\begin{equation}
M_{CT}^{\rm max}[m^2(v)] = \frac{m^2(v)}{m(\delta)} + \frac{m^2(\delta)-m^2(\alpha)}{m(\delta)}.
\label{eqn3}
\end{equation}
Consequently a measurement of the gradient and intercept of the linear
function describing the dependence of $\mctmax$ on $m^2(v)$ allows
both $m(\delta)$ and $m(\alpha)$ to be measured independently.

Despite the simplicity and ease-of-use of the contransverse mass
technique, it suffers from two principle draw-backs
\cite{Tovey:2008ui}. The first is that $\mct$ is not invariant under
Lorentz boosts of the $\delta_1\delta_2$ centre-of-momentum (CoM)
frame in the laboratory transverse plane. Consequently if the
$\delta_1\delta_2$ system recoils in the transverse plane against
upstream object(s) such as ISR jets then the value of $\mct$
calculated in the laboratory frame is not in general equal to that
calculated in the $\delta_1\delta_2$ CoM frame. $\mct$ values can be
generated which are greater than $\mctmax$ and as a result the $\mct$
end-point can be smeared (see e.g. Figure 2 in
Ref.~\cite{Tovey:2008ui}). The second draw-back is apparent when
attempting to measure $m(\delta)$ and $m(\alpha)$ independently in
events with non-zero visible masses $m(v_i)$ using
Eqn.~(\ref{eqn3}). The requirement $m(v_1)=m(v_2)$ can significantly
reduce the available event statistics and require the accumulation of
very large integrated luminosity, even for channels with relatively
large $\sigma.BR$. This problem is illustrated clearly in Figure 3 in
Ref.~\cite{Tovey:2008ui}. This paper will seek to address these two
problems and demonstrate the utility of the $\mct$ technique through
two case-studies. In the process we shall identify a further problem
with using Eqn.~(\ref{eqn3}) to measure masses independently, but
develop an alternative strategy for two-step sequential two-body decay
chains combining $\mct$ end-point measurements with conventional
invariant mass end-point constraints. We shall also investigate the
use of the transverse boost dependence of $\mctmax$ to measure masses
independently, but find that this technique suffers from similar
problems.

The structure of the paper is as follows. Section~\ref{sec2} will
study the transformation properties of $\mct$ under contra-linear and
co-linear Lorentz boosts of $\delta_i$, leading to the development of
a procedure for correcting $\mct$ for co-linear
boosts. Section~\ref{sec2a} will discuss the shape of the resulting
$\mct$ distributions. Section~\ref{sec3} will propose a new method for
maximising the available event statistics when measuring $m(\delta)$
and $m(\alpha)$ independently with Eqn.~(\ref{eqn3}) by removing the
$m(v_1)=m(v_2)$ requirement. This section will also develop a
technique through which $m(\delta)$ and $m(\alpha)$ can in principle
be measured independently by using the transverse boost dependence of
$\mctmax$. Section~\ref{sec4} will investigate these techniques with
LHC case studies measuring the masses of the top quark, $W$ and
neutrino with fully-leptonic $t\bar{t}$ events, and the masses of SUSY
particles decaying to a similar final state. Section~\ref{sec5} will
conclude.

%% file: theory.tex
\section{Transformation properties of $\mct$}\label{sec2}

\subsection{Equal magnitude contra-linear boosts}\label{subsec2.1}

It is instructive to consider first the transformation properties of
$\mct$ under contra-linear equal magnitude boosts, in which $\delta_1$
and $\delta_2$ move in opposite directions with equal momentum. $\mct$
is derived from the quantity $M_C$ given by
\begin{eqnarray}
\label{eqn4}
M_C^2(v_1,v_2) & \equiv & [E(v_1)+E(v_2)]^2 - [{\bf p}(v_1)-{\bf
p}(v_2)]^2 \cr & = & m^2(v_1) + m^2(v_2) + 2[E(v_1)E(v_2)+{\bf p}(v_1)
\cdot {\bf p}(v_2)],
\end{eqnarray}
and this was shown in Ref.~\cite{Tovey:2008ui} to be invariant under
such boosts. By contrast $\mct$ is not in general invariant under such
boosts, however the position of the $\mct$ end-point, $\mctmax$, {\it
is}. The reason for this can be understood by observing
that one can rewrite Eqn.~(\ref{eqn4}) in the following form in the
$\delta_1\delta_2$ CoM frame:
\begin{equation}
\label{eqn5}
M_C^2(v_1,v_2)=m^2(v_1) + m^2(v_2) + 2[E_T(v_1)E_T(v_2)\cosh\Sigma\eta(v_i)+{\bf p_T}(v_1) \cdot {\bf p_T}(v_2)].
\end{equation}
Comparing with Eqn.~(\ref{eqn1}) and noting that
$\cosh\Sigma\eta(v_i)\geq 1$ this shows that $\mct \leq \mc$, with
equality when $\Sigma\eta(v_i)=0$. Now $\mc$, like $\mct$, is bounded
from above by $\mctmax$ and so one finds finally that $\mct \leq \mc
\leq \mctmax$. It is interesting to note additionally that $\mc$
equals $\mctmax$ when $v_1$ and $v_2$ are co-linear in the $\delta_i$
rest frames and hence the necessary and sufficient criteria for
$\mct=\mctmax$ are that $\Delta \eta(v_1,v_2)=0$ and
$\Delta\phi(v_1,v_2)=0$ in the $\delta_i$ rest frames while
$\Sigma\eta(v_i)=0$ in the $\delta_1\delta_2$ CoM frame.

A similar argument applies when the transverse mass $\mtrans$
undergoes co-linear equal magnitude boosts, and the result is similar,
namely that $\mtrans$ is not invariant under arbitrary transverse
boosts but nevertheless possesses a boost-invariant end-point. It is
interesting to note that in this case the necessary and sufficient
conditions for $\mtrans=\mtransmax$ are in some sense the complement
of those in the $\mct$ case: here $\Sigma\eta(v_i)=0$ and
$\Delta\phi(v_1,v_2)=\pi$ in the rest frame(s) of the parent
particle(s) while $\Delta \eta(v_1,v_2)=0$ in the event CoM frame. Of
course if $v_1$ and $v_2$ are the sole products of the decay of the
same parent then the first and second criteria are generally satisfied
through conservation of momentum.

\subsection{Equal magnitude co-linear boosts}\label{subsec2.2}

The contransverse mass is invariant by construction under co-linear
equal magnitude boosts of $\delta_1$ and $\delta_2$ in the beam
($\hat{z}$) direction, by virtue of its dependence purely on
transverse quantities. Consequently $\mctmax$ is similarly
invariant. In the presence of co-linear equal magnitude boosts in the
CoM transverse plane however, equivalent to a single global transverse
boost, the values of both $\mct$ and $\mctmax$ can depend on the
magnitude and direction of the boost. For instance, when
$m(v_1)=m(v_2)=m(v)$ Eqn.~(\ref{eqn3}) becomes:
\begin{equation}
\label{eqn12c}
M_{CT}^{\rm max}[m^2(v),p_b] = 2\big(rp_0+E_0\sqrt{1+r^2}\big),
\end{equation}
where $r\equiv p_b/2m(\delta)$, $p_b$ is the net transverse momentum
of upstream objects (ISR jets etc.) generating the boost,
\begin{equation}
E_0 \equiv \frac{m^2(\delta)-m^2(\alpha)+m^2(v)}{2m(\delta)},
\end{equation}
and $p_0\equiv\sqrt{E_0^2-m^2(v)}$.

With sufficient statistics one might hope to use Eqn.~(\ref{eqn12c}) to
measure $m(\delta)$ and $m(\alpha)$ separately by measuring $\mctmax$
as a function of $p_b$. This possibility is considered further in
Section~\ref{sec3}. With limited statistics however it would be useful
to be able to transform $\mct$ such that its value always lies below
the $\mct$ end-point given by Eqn.~(\ref{eqn3}). In this case one
sacrifices the ability to measure $m(\delta)$ and $m(\alpha)$
independently using the $p_b$ dependence of $\mctmax$ in order to
maximise statistics near the $p_b=0$ $\mct$ end-point, while limiting
smearing beyond the end-point due to integration over $p_b$.

The approach we shall take will involve boosting the four-momenta of
the visible decay products $v_i$ back into the $\delta_1\delta_2$ CoM
frame with boost factor $\beta$, prior to calculating $\mct$. If we
know neither the sign nor the magnitude of $\beta$ then the minimum
value of $\mct$ that we can obtain by varying the assumed value of
$\beta$ is the one-dimensional analogue of $\mct$ given by $\mcy$
defined by
\begin{eqnarray}
\label{eqn6}
M_{Cy}^2(v_1,v_2) & \equiv & [E_y(v_1)+E_y(v_2)]^2 - [p_y(v_1)-p_y(v_2)]^2\cr
& = & m^2(v_1) + m^2(v_2) + 2[E_y(v_1)E_y(v_2)+p_y(v_1)p_y(v_2)],
\end{eqnarray}
where $E_y^2(v_i) \equiv p_y^2(v_i) + m^2(v_i)$ and we have assumed
that the boost lies in the $\pm\hat{x}$ direction. $\mcy$ is invariant
under arbitrary boosts in the $\hat{x}-\hat{z}$ plane for the same
reason that $\mct$ is invariant under boosts in the $\hat{z}$
direction. It represents a conservative lower bound on the value
$\mcto$ of $\mct$ measured in the $\delta_1\delta_2$ CoM frame. The
criterion for $\mct$ to equal $\mcy$ in any given frame, and hence for
$\mct$ to be minimised in that frame, is that $A_x=0$, where $A_x$ is
defined by
\begin{equation}
\label{eqn7}
A_x \equiv p_x(v_1)E_y(v_2)+p_x(v_2)E_y(v_1).
\end{equation}

As an aside, it is interesting to consider at this point the
possibility of using $M_C$ defined by Eqn.~(\ref{eqn4}), rather than
$\mct$, and attempting to perform a correction for longitudinal boosts
along the beam direction. In this case we know neither the sign nor
the magnitude of the $z$-boost and so by the above argument the
appropriate quantity to use is the two-dimensional analogue of
$M_{C}$, which is just $\mct$ as we have been using already. The
criterion for $M_C$ to equal $\mct$ is by analogy with
Eqn.~(\ref{eqn7}) 
\begin{equation}
p_z(v_1)E_T(v_2)+p_z(v_2)E_T(v_1)=0,
\end{equation}
which is equivalent to setting $\Sigma\eta(v_i)=0$ as required by
Eqn.~(\ref{eqn5}).

Now in fact we {\it do} know the sign of the required boost, because
we know the direction in the transverse plane of the upstream
momentum. Defining this direction to be the $+\hat{x}$ direction we
need to boost $v_1$ and $v_2$ in this same direction (i.e. use
$\beta\geq0$) to correct for the original boost of the
$\delta_1\delta_2$ CoM frame, which must have been in the $-\hat{x}$
direction. Such a $+\hat{x}$ boost monotonically increases $p_x(v_i)$
and hence monotonically increases the transformed value of $A_x$
towards $+\infty$ as $\beta\rightarrow +1$. Let us define now $\axl$
and $\mctl$ to be the values of $A_x$ and $\mct$ measured in the lab
frame and $\axp$ and $\mctlc$ to be the equivalent values obtained
after boosting $v_1$ and $v_2$. If $\axl\geq0$ then $\mctlc$ increases
monotonically from $\mctl$ towards $+\infty$ as $\beta\rightarrow +1$
(see Figures~\ref{fig0} and~\ref{fig0a}). Consequently in this case
the least conservative lower bound on $\mcto$ we can obtain is
$\mctlc=\mctl$. If on the other hand $\axl$ is negative then as
$\beta$ and $\axp$ increase $\mctlc$ first decreases from $\mctl$
towards its minimum value of $\mcy$ (at $\axp=0$) before increasing
again towards $+\infty$ (see Figures~\ref{fig0} and~\ref{fig0a}). In
this case, without further information, the best we can do is set
$\mctlc=\mcy$.
\FIGURE[ht]{
\epsfig{file=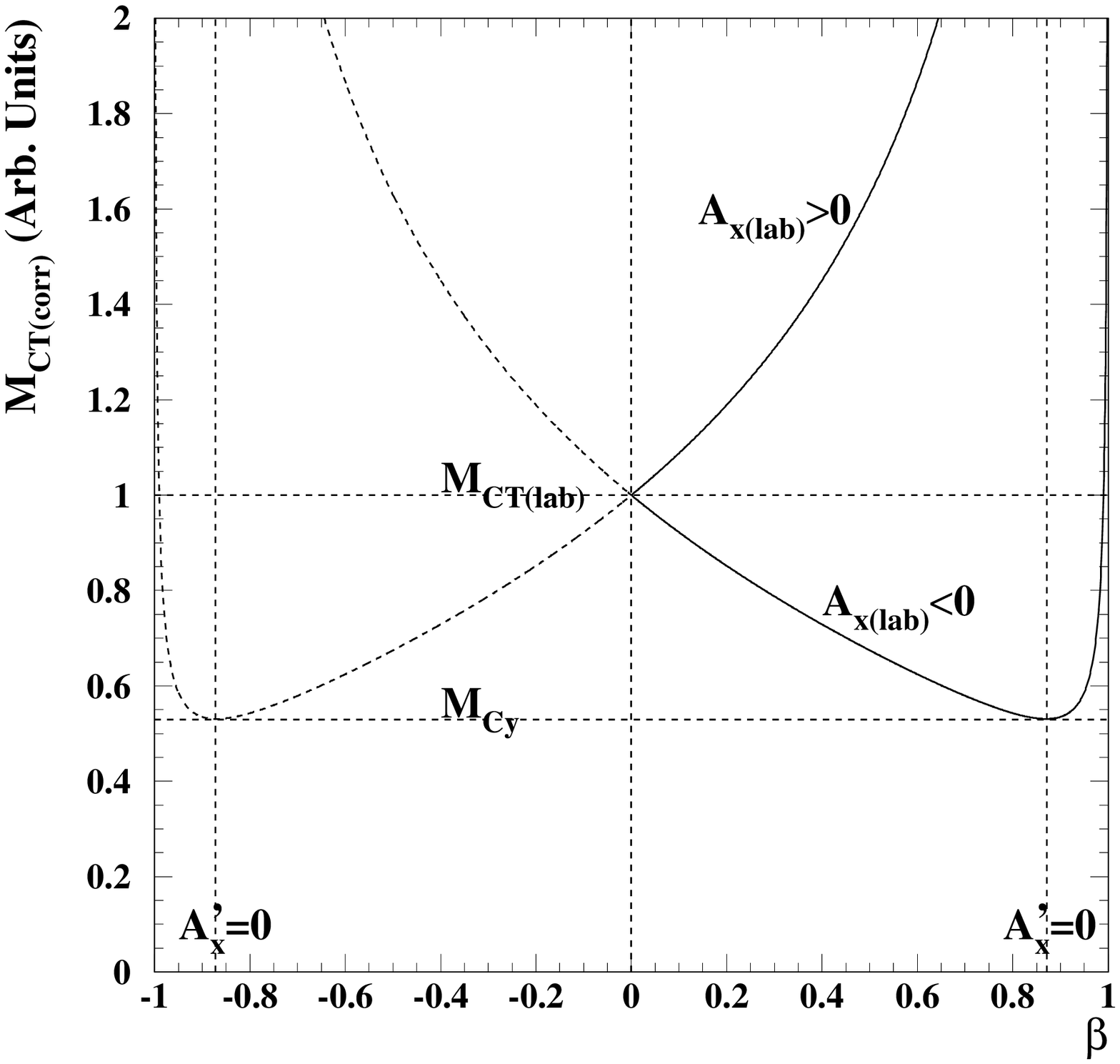,height=4.0in}
\caption{\label{fig0} Schematic diagram showing the dependence of
$\mctlc$ on the boost factor $\beta$ used in the boost
correction. Cases with $\axl>0$ and $\axl<0$ are shown. When
$\beta\geq0$ and $\axl\geq0$ the minimum value of $\mctlc$ occurs when
$\beta=0$ and hence $\mctlc=\mctl$. If $\beta\geq0$ and $\axl<0$ the
minimum value of $\mctlc$ is $\mcy$.} }
\FIGURE[ht]{ \epsfig{file=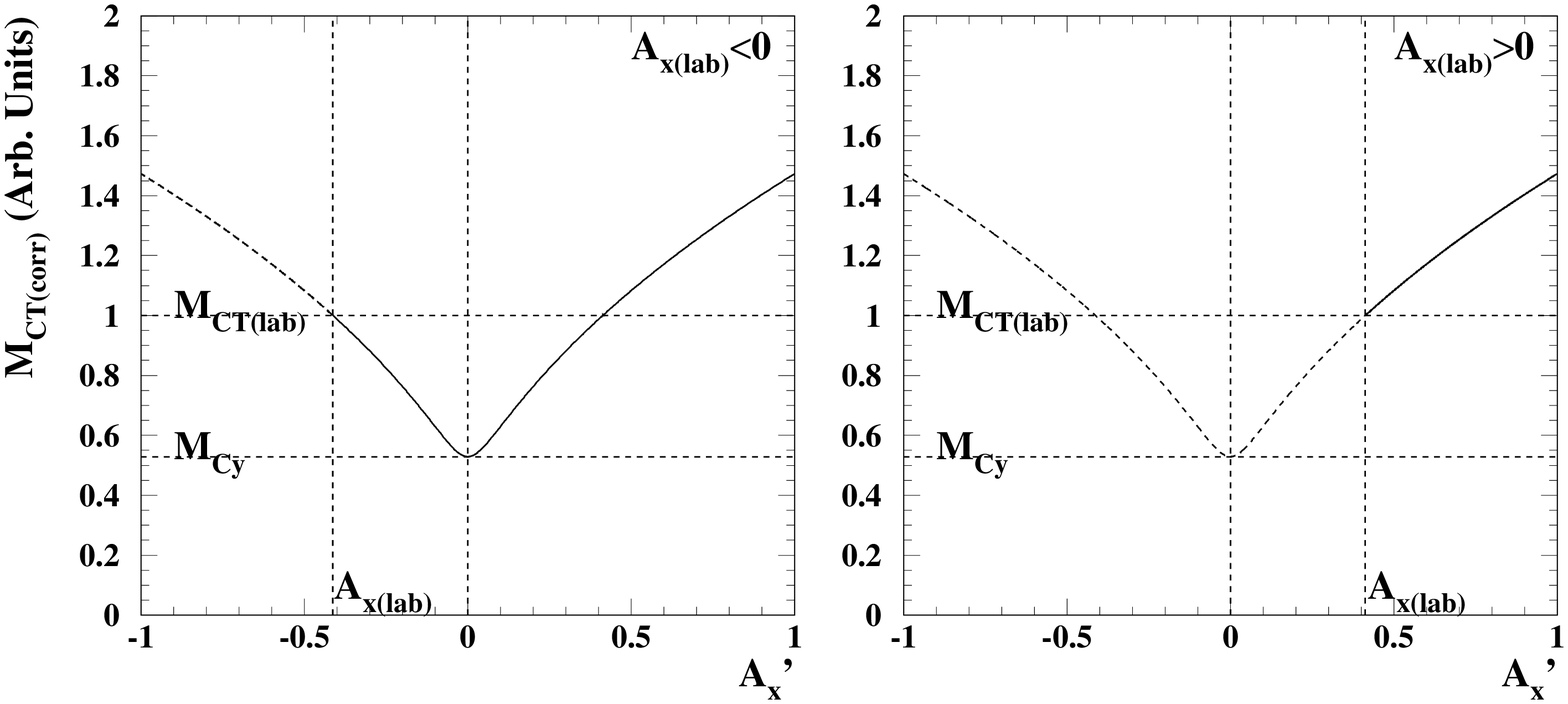,height=3.0in}
\caption{\label{fig0a} Schematic diagram showing the dependence of
$\mctlc$ on $\axp$. If $\axl<0$ (left-hand figure) then as $\axp$
increases from $\axl$, $\mctlc$ passes through its minimum at
$\axp=0$, while if $\axl>0$ (right-hand figure) it does not.} }

Fortunately however we have not yet exhausted the possibilities for
boost correction. Observe first that when boosting $v_1$ and $v_2$ the
boost factor is given by $\beta=p_b/\eog$ where $\eog$ is the assumed
value of the energy $E_{\delta\delta}$ of the $\delta_1\delta_2$ CoM frame
in the lab frame. Consequently increasing the value of $\beta$ is
equivalent to decreasing $\eog$, and vice versa. Hence if $\mctlc$
increases (decreases) monotonically with increasing $\eog$ and we set
$\eog$ to a value less than (greater than) $E_{\delta\delta}$, the value of
$\mctlc$ we obtain provides a conservative lower bound on $\mcto$. Now
if $\axl\geq0$ then $\mctlc$ always increases with increasing $\beta$
(see Figure~\ref{fig0}) and hence it decreases with increasing $\eog$
(see Figure~\ref{fig1} -- upper curve). In this case we should set
$\eog$ to the upper bound on $E_{\delta\delta}$, boost $v_1$ and $v_2$, and obtain
a conservative lower bound on $\mcto$ from the value of $\mctlc$ in
this frame. If $\axl<0$ the situation is more complicated (see
Figure~\ref{fig1} -- lower curve). In this case, if $\axp<0$ after
boosting with $\eog$ set to both the upper and lower bounds on $E_{\delta\delta}$
then the least conservative lower bound on $\mcto$ is given by the
value of $\mctlc$ with $\eog$ set to the lower bound on
$E_{\delta\delta}$. Conversely if $\axp\geq0$ in both these cases then the least
conservative lower bound on $\mcto$ is given by the value of $\mctlc$
with $\eog$ set to the upper bound on $E_{\delta\delta}$. If $\axp\geq0$ with
$\eog$ set to the lower bound on $E_{\delta\delta}$ but $\axp<0$ with $\eog$ set to
the upper bound on $E_{\delta\delta}$ then $\mcto$ could be as low as $\mcy$ and so
this should be used as the least conservative lower bound on $\mcto$.
\FIGURE[ht]{ \epsfig{file=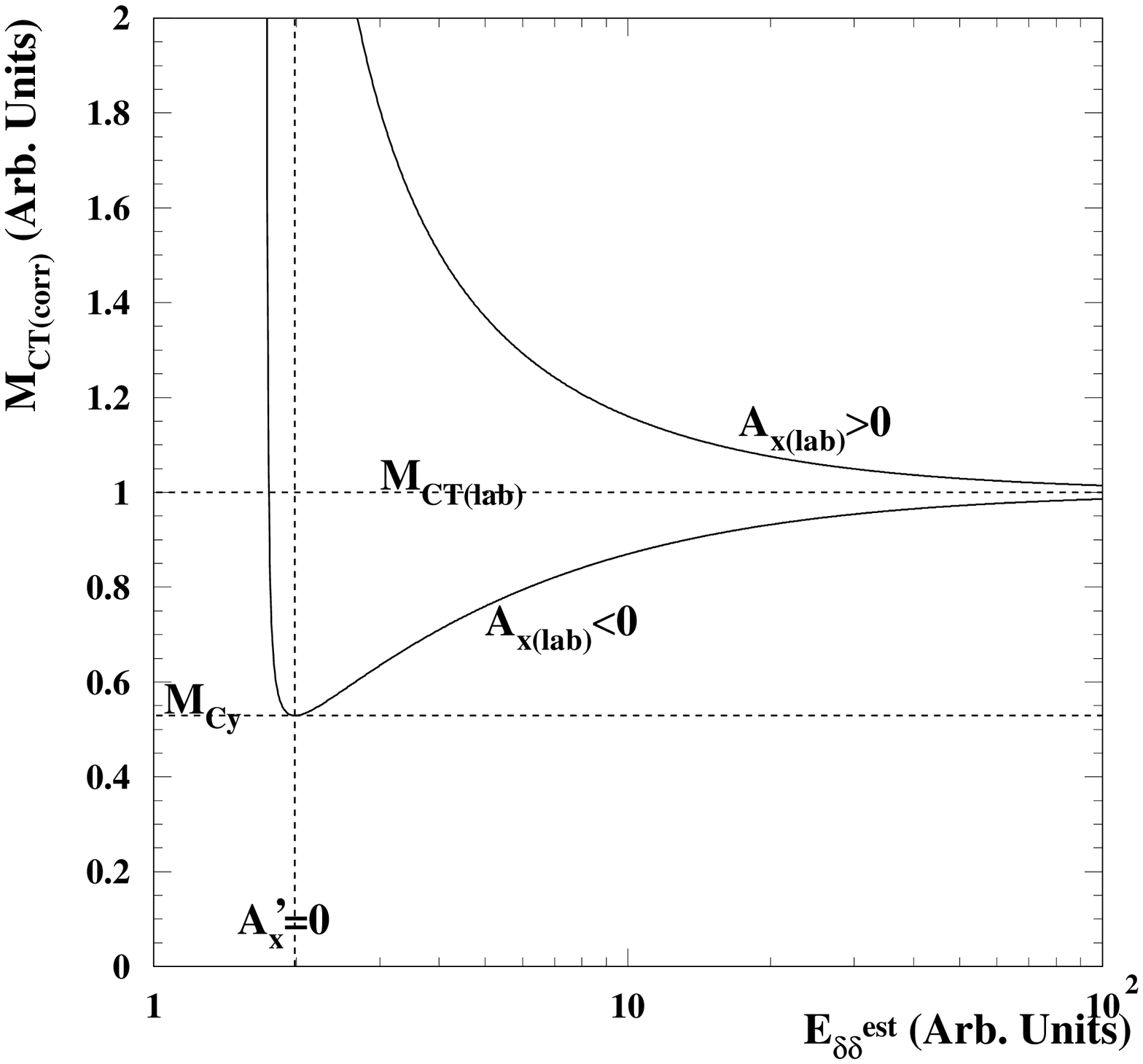,height=4.0in}
\caption{\label{fig1} Schematic diagram showing the dependence of
$\mctlc$ on $\eog$, the estimated value of $E_{\delta\delta}$ used in the
boost. Cases with $\axl>0$ and $\axl<0$ are shown. When $\axl\geq0$ or
$\axp> 0$ $\mctl$ decreases with increasing $\eog$. If however
$\axl<0$ and $\axp<0$ then $\mctl$ increases with increasing $\eog$. }}

In fact we can indeed obtain both upper and lower bounds on $E_{\delta\delta}$. An
upper bound is provided by the proton-proton centre of mass energy
$E_{\rm cm}$ while the total visible energy $\ehat$ of the decay
products provides a lower bound. This latter quantity is calculated by
summing the energies of the visible decay products with the net
transverse momentum of the invisible decay products\footnote{If a
lower bound $m_{\rm lo}(\alpha)$ on the masses of the individual
invisible decay products can be assumed then conservatively one can
use $\sqrt{(E_T^{\rm miss})^2+4m_{\rm lo}^2(\alpha)}$ in $\ehat$
instead of $E_T^{\rm miss}$ to obtain an improved bound on $E_{\delta\delta}$.}
given by $E_T^{\rm miss}$. $E_T^{\rm miss}$ equals the total energy of
the invisible decay products only when these are massless, co-linear,
and moving in the transverse plane, and so in general $\ehat\leq
E_{\delta\delta}$. Below we shall denote values of $\axp$ obtained with $\eog$ set
to $E_{\rm cm}$ or $\ehat$ as respectively $\axecm$ and $\axmthard$.

Let us now summarise the procedure we have developed for correcting
$\mct$ for the effects of co-linear equal magnitude boosts of
$\delta_1$ and $\delta_2$ in the transverse plane\footnote{f77, C++
and {\tt ROOT} code implementing this boost-correction procedure can
be downloaded from {\tt http://projects.hepforge.org/mctlib}.}. First
calculate $\axl$ and $\axecm$ using Eqn.~(\ref{eqn7}), the latter by
boosting $v_1$ and $v_2$ with $\beta=p_b/E_{\rm cm}$. If $\axl\geq 0$
or $\axecm\geq0$ then one should set $\mctlc$ to the boosted value of
$\mct$ obtained with $\beta=p_b/E_{\rm cm}$. If neither of these
criteria are satisfied then one should next evaluate $\ehat$ and boost
$v_1$ and $v_2$ with $\beta=p_b/\ehat$. Now evaluate $\axmthard$ using
Eqn.~(\ref{eqn7}). If $\axmthard<0$ then one should set $\mctlc$ to
the value of $\mct$ in this boosted frame. If however $\axmthard\geq0$
then one should set $\mctlc=\mcy$. An example of the effect of this
boost correction procedure is shown in Fig.~\ref{fig2} for the SUSY
events considered in Ref.~\cite{Tovey:2008ui}.  
\FIGURE[ht]{
\epsfig{file=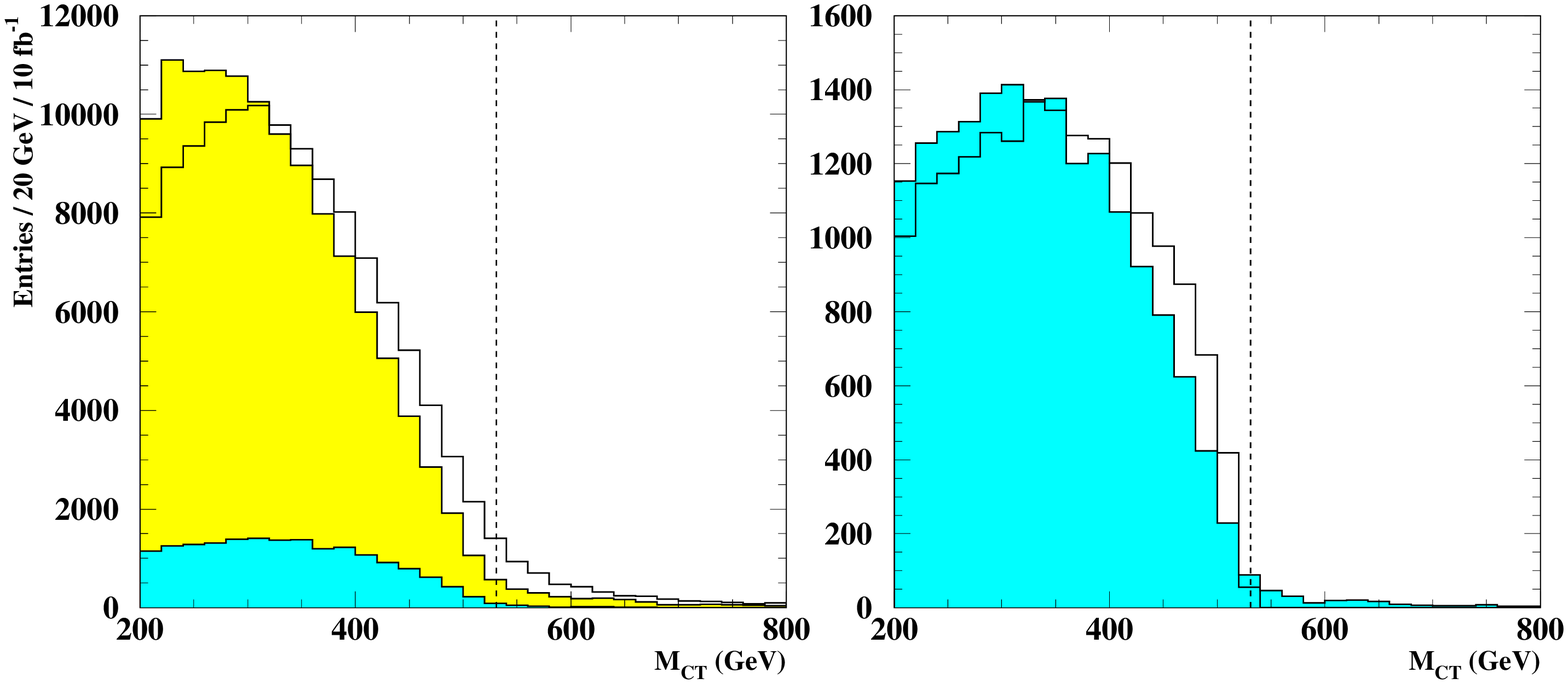,height=3.0in}
\caption{\label{fig2} $\mct$ distributions of SUSY events containing
at least two jets for the SPS1a benchmark SUSY model considered in
Ref.~\cite{Tovey:2008ui}. In the left-hand figure the open histogram
shows the $\mct$ distribution with no $p_b$ cut and no correction
applied. This diagram should be compared with Figure 2 of
Ref.~\cite{Tovey:2008ui} showing the evolution of the uncorrected
$\mct$ end-point as a function of the cut on $p_b$. The light (yellow)
histogram shows the $\mct$ distribution of the same events after the
co-linear boost correction described in the text has been applied to
the $\mct$ values, assuming that all additional jets contribute to the
upstream momentum $p_b$. The medium (cyan) histogram shows the same
distribution for $\sqr\sqr$ pair-production events.  In the right-hand
figure the latter distribution is plotted on an expanded scale. The
open histogram shows the parton-level $\mct$ distribution for the same
events.  The end-point from $\sqr\sqr$ pair-production is expected at
531 GeV (denoted by a vertical dashed line in both figures). } }

%% file: indmass.tex
\section{The shape of the $\mct$ distribution}\label{sec2a}

The differences between $\mtrans$ and $\mct$ identified in
Section~\ref{subsec2.1} affect the shapes of the distributions of
these quantities. As is well-known, $\mtrans$ possesses a Jacobian
peak at $\mtrans=\mtransmax$ when $v_1$ and $v_2$ are the sole
products of the decay of the same parent. Physically this peak arises
because near the end-point all kinematic configurations with different
$\eta(v_i)$ generate the same value of $\mtrans$, in other words
$\mtrans$ becomes independent of the kinematics of $v_1$ and $v_2$.

Turning now to $\mct$, let us consider first the special case where
the $\delta_1\delta_2$ system is not boosted in the laboratory
transverse plane and no boost correction is applied. When $\mct$ is
calculated for the visible decay products of the $\delta_1\delta_2$
system the extra degrees of freedom resulting from the independent
motion of $v_1$ and $v_2$ generate a significantly different shape of
distribution. Near the end-point at $\mctmax$ only events in which
both $v_1$ and $v_2$ move in the transverse plane can contribute to
the distribution. The small probability of this configuration (because
$v_1$ and $v_2$ are uncorrelated) cancels the large probability
generated by the Jacobian transformation, resulting in an end-point
which tends asymptotically in the absence of boosts to
\begin{equation}
\label{eqn100}
{\rm P}(\mct) {\rm \mbox{ } d}\mct=A\sqrt{(\mctmax)^2-\mct^2} {\rm \mbox{ } d}\mct, 
\end{equation}
where $A$ is a constant. Typical $\mct$ distributions in the absence of
boosts, displaying this end-point, are shown in
Figure~\ref{fig-mctschem}.
\FIGURE[ht]{
\epsfig{file=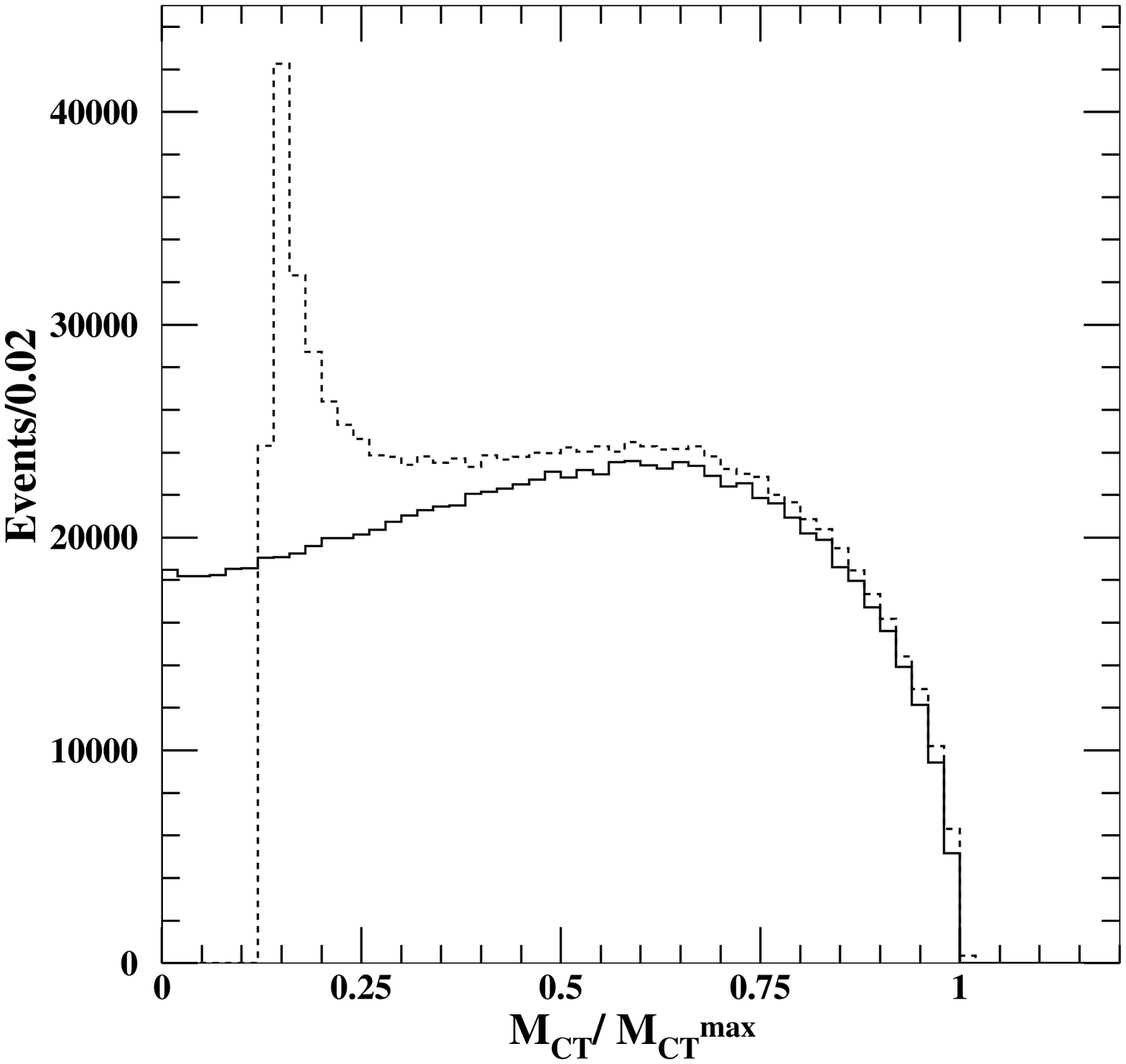,height=4.0in}
\caption{\label{fig-mctschem} Typical $\mct$ distributions in the
absence of boosts for massless visible particles (full histogram) and
massive visible particles of total mass $0.13\mctmax$ (dashed
histogram). The $x$-axis has been scaled such that the end-point lies
at $\mct/\mctmax=1$.}}

Despite this cancellation of the peak at $\mctmax$ the $\mct$
distribution can still possess a Jacobian peak. The peak occurs however at
the lower limit of the distribution, where $\mct = m(v_1)+m(v_2) \equiv
\mctmin$ (see Eqn.~(\ref{eqn1})). The distribution tends
asymptotically in the absence of boosts to
\begin{equation}
\label{eqn101}
{\rm P}(\mct) {\rm \mbox{ } d}\mct=B\frac{\mct}{\sqrt{\mct^2-(\mctmin)^2}} {\rm \mbox{ } d}\mct, 
\end{equation}
where $B$ is a constant. This is very similar to the functional form
of the $\mtrans$ Jacobian peak, although in this case it is reversed
such that the distribution is real above the peak rather than below
it. Physically the peak occurs because when $\mct \sim \mctmin$ the
value of $\mct$ becomes independent of the kinematics of $v_1$ and
$v_2$, because the $m(v_i)$ terms in Eqn.~(\ref{eqn1}) dominate. Hence
all kinematic configurations generate similar values of $\mct$. An
example of such a peak can be seen in the dashed histogram in
Figure~\ref{fig-mctschem}. Note that when $\mctmin=0$ the numerator
and denominator in Eqn.~(\ref{eqn101}) cancel leaving a uniform
distribution (see e.g. full histogram in Figure~\ref{fig-mctschem}).

Consider now the general case where the $\delta_1\delta_2$ system has
been boosted in the laboratory transverse plane, and the boost
correction procedure discussed in Section~\ref{subsec2.2} has been
applied. In this case there can be a further enhancement of the
population of events at $\mct=\mctmin$. If $\axl<0$ and $\axmthard>0$
then the boost-corrected value of $\mct$ is given by $\mcy$ from
Eqn.~(\ref{eqn6}). If the tranverse momenta of the two visible
particles under consideration are bisected by the boost direction
however, then $\mcy$ can take very small values, even if the
transverse momenta of the $v_i$ particles are relatively large. This
effect is particularly striking when $m(v_1)=m(v_2)=\mctmin=0$, in
which case it is straight-forward to see from Eqn.~(\ref{eqn6}) that
$\mcy=0$. We shall refer below to the resulting peak at $\mct=\mctmin$
as the `$\mct=\mcy$' peak.

Specific examples of the peaks and end-points discussed in this
section can be seen in Section~\ref{sec4} below.

\section{Measuring $m(\delta)$ and $m(\alpha)$ independently}\label{sec3}

The dependence of $\mctmax$ on $m(v_i)$, noted in
Ref.~\cite{Tovey:2008ui}, and on $p_b$, noted in
Section~\ref{subsec2.2}, provides potential techniques for measuring
$m(\delta)$ and $m(\alpha)$ independently. In this section we develop
these techniques in more detail.

\subsection{Using the $m(v_i)$ dependence of $\mctmax$}\label{subsec3.1}

As remarked in Section~\ref{sec1}, when attempting to measure
$m(\delta)$ and $m(\alpha)$ independently using Eqn.~(\ref{eqn3}) the
requirement $m(v_1)=m(v_2)=m(v)$ reduces significantly the event
selection efficiency. One can consider ameliorating this problem by
removing this mass equality requirement and considering the dependence
of the resulting $\mct$ end-point on both $m^2(v_1)$ and
$m^2(v_2)$. This is given by
\begin{multline}
\label{eqn12}
\Big(\mctmax[m^2(v_1),m^2(v_2)]\Big)^2 = m^2(v_1) + m^2(v_2) + 2\Big( E_0(v_1)E_0(v_2) \\ + \sqrt{\big[E_0^2(v_1)-m^2(v_1)\big]\big[E_0^2(v_2)-m^2(v_2)\big] }\Big),
\end{multline}
where 
\begin{equation}
\label{eqn13}
E_0(v_i) \equiv \frac{m^2(\delta)-m^2(\alpha)+m^2(v_i)}{2m(\delta)}.
\end{equation}
In this case all events passing background rejection cuts are used,
however the implicit requirement of binning in both $m^2(v_1)$ and
$m^2(v_2)$ to measure $\mctmax[m^2(v_1),m^2(v_2)]$ still limits the
available event statistics in each end-point measurement (modulo the
symmetry under interchange of $v_1$ and $v_2$ of
$\mctmax[m^2(v_1),m^2(v_2)]$). As an aside, Eqn.~(\ref{eqn12})
provides a link between the `stransverse mass' $M_{T2}(\chi)$
\cite{Lester:1999tx} and $\mct$. This is discussed in more detail in
Appendix A.

An alternative to using Eqn.~(\ref{eqn12}) for measuring $m(\delta)$
and $m(\alpha)$ independently involves observing that
$\mctmax[m^2(v)]$ is linearly dependent on $m^2(v)$ in
Eqn.~(\ref{eqn3}) and hence that
\begin{equation}
\label{eqn14}
\mctmax[\mmax^2] = \max\left(\mctmax[m^2(v_1)],\mctmax[m^2(v_2)]\right),
\end{equation}
where,
\begin{equation}
\label{eqn15}
\mmax \equiv \max\left(m(v_1),m(v_2)\right).
\end{equation}
Then we can make use of the following inequality:
\begin{equation}
\label{eqn16}
\mctmax[m^2(v_1),m^2(v_2)] \le \mctmax[\mmax^2],
\end{equation}
to find that 
\begin{equation}
\label{eqn17}
\mct(v_1,v_2) \le \mctmax[m^2(v_1),m^2(v_2)] \le \mctmax[\mmax^2].
\end{equation}
Consequently if the two-dimensional distribution of event
$\mct(v_1,v_2)$ values versus event $\mmax^2$ values is plotted, for
all events passing background rejection cuts, the distribution will
display an $\mct$ end-point dependence on $\mmax^2$ given by:
\begin{equation}
\label{eqn18}
\mctmax[\mmax^2] = \frac{\mmax^2}{m(\delta)} + \frac{m^2(\delta)-m^2(\alpha)}{m(\delta)}.
\end{equation}
Hence $m(\delta)$ and $m(\alpha)$ may be obtained by measuring the
gradient and intercept of the end-point dependence on $\mmax^2$ in a
similar manner to the existing technique using Eqn.~(\ref{eqn3}). Now
however all events passing the background rejection cuts can be used
rather than just a small subset. 

It should be noted here that although this technique is sound from a
theoretical point-of-view, the uneven distribution of events in the
$\mct(v_1,v_2)$ versus $\mmax^2$ plane can cause difficulty when
attempting to use it in practice. This is discussed further in
Section~\ref{sec4}.

\subsection{Using the $p_b$ dependence of $\mctmax$}\label{subsec3.2}

In Section~\ref{subsec2.2} we developed a procedure for correcting
$\mct(v_1,v_2)$ such that it is always bounded from above by the
expression for $\mctmax$ obtained when the upstream momentum
$p_b=0$. Given sufficient statistics however an alternative procedure
would involve binning the non-boost-corrected value of $\mct(v_1,v_2)$
in $p_b$ and measuring $m(\delta)$ and $m(\alpha)$ independently from
the dependence of $\mctmax$ on $p_b$. When $m(v_1)=m(v_2)=m(v)$ this
dependence is given by Eqn.~(\ref{eqn12c}), however we can also obtain
a general expression valid even when $m(v_1)\neq m(v_2)$, which is
\begin{multline}
\label{eqn12a}
\Big(\mctmax[m^2(v_1),m^2(v_2),p_b]\Big)^2 = \Big(\mctmax[m^2(v_1),m^2(v_2)]\Big)^2 + \\ 4r^2 \Big(E_0(v_1)E_0(v_2)+p_0(v_1)p_0(v_2)+ \frac{1}{r}\sqrt{1+r^2}\big[p_0(v_1)E_0(v_2)+p_0(v_2)E_0(v_1)\big]\Big),
\end{multline}
where $\mctmax[m^2(v_1),m^2(v_2)]$ is obtained from
Eqn.~(\ref{eqn12}), $r\equiv p_b/2m(\delta)$, $E_0(v_i)$ is given by
Eqn.~(\ref{eqn13}) and $p_0(v_i)\equiv\sqrt{E_0^2(v_i)-m^2(v_i)}$. In
principle the additional dependence of $\mctmax$ on the angle between
the $p_b$ vector and the net momentum of $v_1$ and $v_2$ in the
transverse plane could also be exploited, although this is not
considered further here.

The advantage of using the $p_b$ dependence of $\mctmax$ rather than
the $m(v_i)$ dependence discussed in Section~\ref{subsec3.1} is that
$m(\delta)$ and $m(\alpha)$ can in principle be measured independently
even when $m(v_1)=m(v_2)=m(v)\simeq 0$, for instance when the $v_i$
particles are jets or leptons. This avoids potential combinatorial
problems inherent in the latter technique. In this special case the
expression for $\mctmax$ becomes
\begin{equation}
\label{eqn12d}
\mctmax[0,0,p_b] = \mctmax[0,0,0]\big(r+\sqrt{1+r^2}\big).
\end{equation}
A toy Monte Carlo example $\mct(v_1,v_2)$ versus $p_b$ distribution
for massless $v_i$ particles is shown in Figure~\ref{fig-pbtheo}
together with the theoretical bound from Eqn.~\ref{eqn12d}.
\FIGURE[ht]{ \epsfig{file=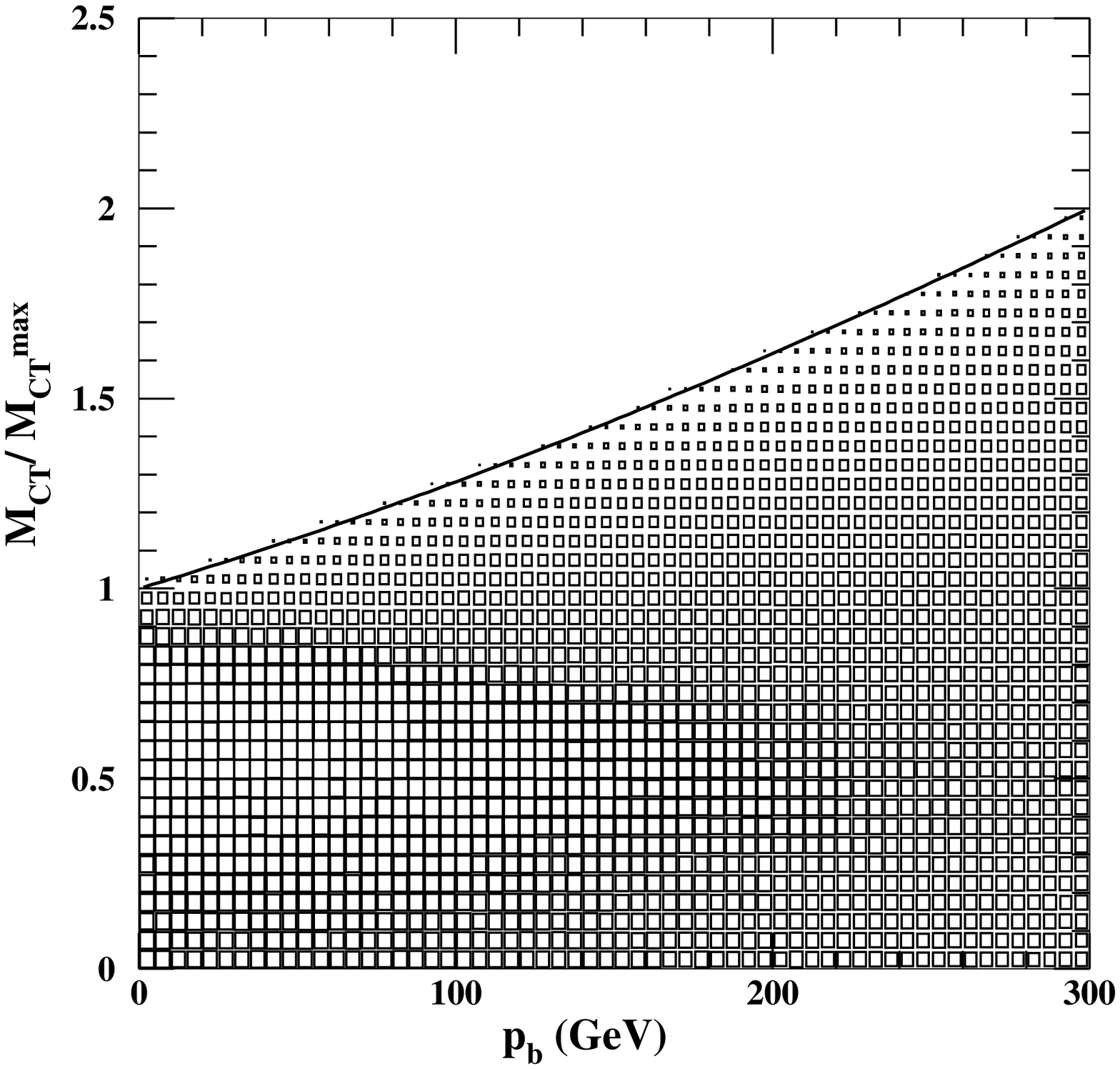,height=4.0in}
\caption{\label{fig-pbtheo} Example $\mct(v_1,v_2)$ versus $p_b$
distribution generated with a toy Monte Carlo showing the dependence
of $\mctmax$ on $p_b$ given by Eqn.~(\ref{eqn12d}). The simulated
events consist of pair-produced $\delta$ particles of mass 200 GeV,
each decaying into a massless visible particle and an invisible
particle $\alpha$ of mass 100 GeV. $p_b$ is evenly distributed in the
range 0--300 GeV. The $\mct(v_1,v_2)$ axis has been normalised to
$\mctmax$ obtained when $p_b=0$. } }

As with the use of the $m(v_i)$ dependence of $\mctmax$ in
Section~\ref{subsec3.1} the above technique is sound from a
theoretical perspective, however as we shall see in Section~\ref{sec4}
the uneven distribution of events in the $\mct(v_1,v_2)$ versus $p_b$
plane can cause difficulty when attempting to use it in practice.

%% file: twostep.tex
\section{Mass measurement for two-step decay chains}\label{sec4}

\subsection{Generic strategy}\label{subsec4.0}

We shall now investigate the use of the boost-corrected contransverse
mass discussed in Section~\ref{subsec2.2} (hereafter referred to
simply as $\mct$) to measure the masses of pair-produced heavy
particles decaying via symmetric two-step sequential two-body decay
chains. As discussed in Ref.~\cite{Burns:2008va} mass measurement with
such chains is non-trivial because they are too short to solve fully
for the masses using invariant mass end-point techniques
\cite{Bachacou:2000zb,Allanach:2000kt,Costanzo:2009mq,Burns:2009zi,Matchev:2009iw}
or the `mass relation' method \cite{Kawagoe:2004rz}. The decay chains
considered can be written in the form:
\begin{equation}
\delta \rightarrow P\beta \rightarrow PQ \alpha,
\label{eqn19}
\end{equation}
where $\delta$, $\beta$ and $\alpha$ are generic massive particles,
$P$ and $Q$ are generic visible particles (here assumed massless) and
$\alpha$ is invisible. The two chains present in each event can be
seen in diagrammatic form in Figure~\ref{figdecay}, where particles
appearing in the second decay chain are denoted with primed labels. We
assume in the following discussion that particles labeled with the
same letter possess the same mass.

With each event we can construct one pair of invariant mass
observables and three contransverse mass observables from the momenta
of the four observed particles $P$, $Q$, $P'$ and $Q'$. These
observables are:
\begin{itemize}
\item $m(P^{(\prime)},Q^{(\prime)})$: the invariant masses of the
visible products of the two decay chains
\item $\mct(P,P')$: $\mct$ constructed from the momenta of $P$ and $P'$
\item $\mct(Q,Q')$: $\mct$ constructed from the momenta of $Q$ and $Q'$
\item $\mct([PQ],[P'Q'])$: $\mct$ constructed from the momenta of the
aggregate products of each chain $[PQ]$ and $[P'Q']$. 
\end{itemize}
These observables possess kinematic end-points whose positions are
functions of the masses $m(\delta)$, $m(\beta)$ and $m(\alpha)$. The
end-point positions are respectively\footnote{When dealing with
single-step three-body decay chains in which two visible particles and
one invisible particle are produced in each decay
Eqns.~(\ref{eqn20a})--(\ref{eqn20d}) are replaced by $m^{\rm
max}(P,Q)=m(\delta)-m(\alpha)$,
$\mctmax([PQ],[P'Q'])=2[m(\delta)-m(\alpha)]$ and
$\mctmax(P,P')=\mctmax(Q,Q')=[m^2(\delta)-m^2(\alpha)]/m(\delta)$,
although in the last two cases the distributions are strongly
phase-space suppressed near the endpoints.}:
\begin{eqnarray}
m^{\rm max}(P,Q)&=&\frac{\sqrt{[m^2(\delta)-m^2(\beta)]
[m^2(\beta)-m^2(\alpha)]}}{m(\beta)}\equiv k_1, \label{eqn20a}\\
\mctmax(P,P')&=&\frac{m^2(\delta)-m^2(\beta)}{m(\delta)}\equiv k_2,\label{eqn20b}\\
\mctmax(Q,Q')&=&\frac{m^2(\beta)-m^2(\alpha)}{m(\beta)}\equiv k_3,\label{eqn20c}\\
\mctmax([PQ],[P'Q'])&=&\frac{m^2(\delta)-m^2(\beta)}{m(\delta)}
+m(\delta)\left(\frac{m^2(\beta)-m^2(\alpha)}{m^2(\beta)}\right)\equiv k_4,\label{eqn20d}
\end{eqnarray}
where the final relationship is obtained from Eqn.~(\ref{eqn3}) with
$m(v)=m^{\rm max}(P,Q)$. In addition the two-dimensional distribution
of events in the $\mct([PQ],[P'Q'])$ versus $\mmax^2$ plane discussed
in Section~\ref{subsec3.1} can be constructed, providing additional
mass constraints via Eqn.~(\ref{eqn18}). The two-dimensional
distributions of events in the (non-boost-corrected) $\mct(P,P')$
versus $p_b$ and $\mct(Q,Q')$ versus $p_b$ planes discussed in
Section~\ref{subsec3.2} can also in principle be used.
\FIGURE[ht]{
\epsfig{file=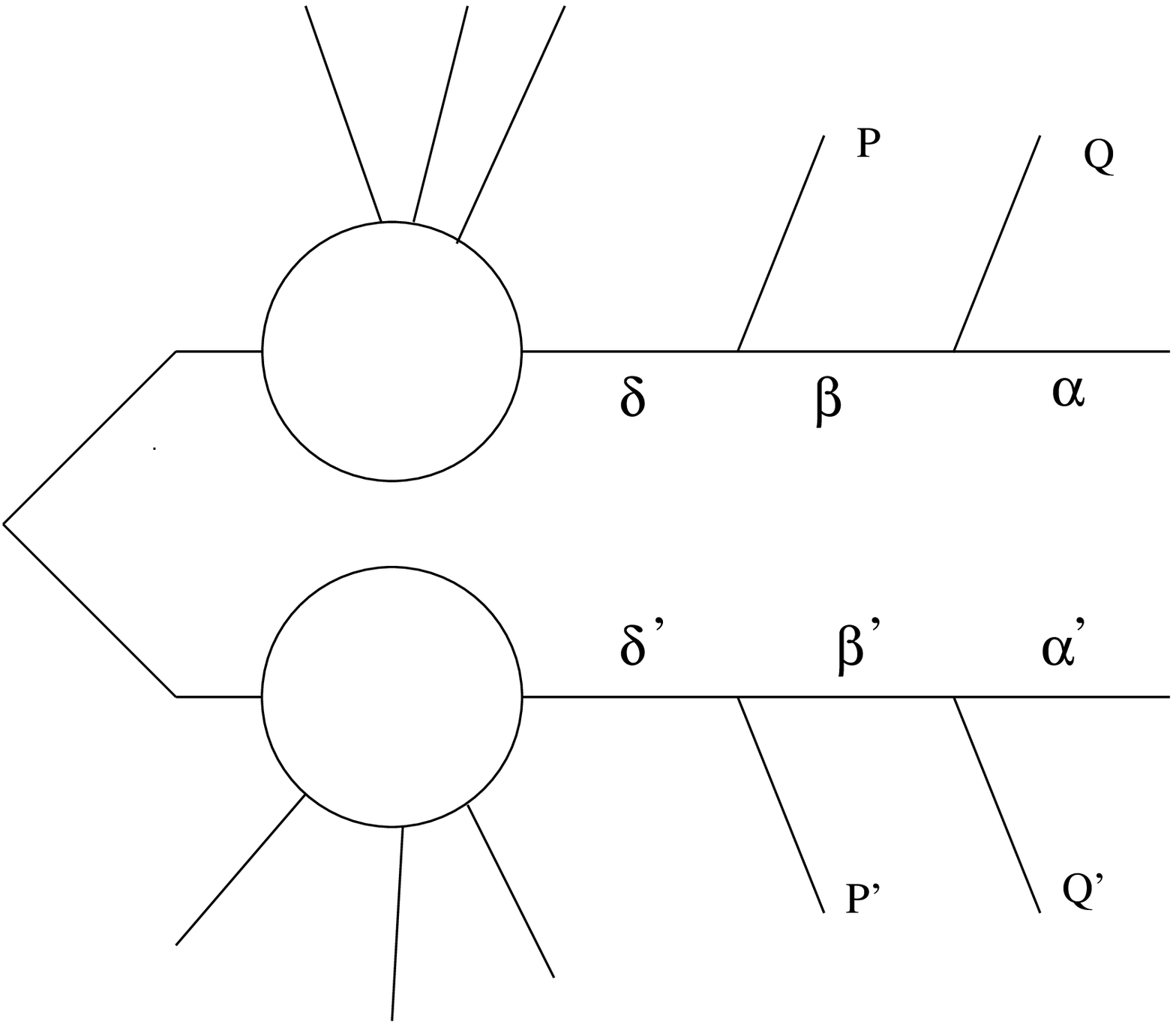,height=4.0in}
\caption{\label{figdecay} Diagrammatic view of the decay chain
described in the text. The mass measurement technique described in the
text is independent of the boost given to the system of interest by
upstream decays or ISR (denoted by circles).}}

Using Eqns.~(\ref{eqn20a}),~(\ref{eqn20b}) and~(\ref{eqn20c}) the mass
of the parent particle $\delta$ can be calculated from:
\begin{equation}
m(\delta)=\frac{k_1^4k_2}{k_1^4-k_2^2k_3^2}.
\label{eq:ma}
\end{equation}
The masses of $\beta$ and $\alpha$ can then be obtained by simple
substitution into Eqns.~(\ref{eqn20b}) and~(\ref{eqn20c})
respectively. The constraints on the masses provided by
Eqn.~(\ref{eqn18}), Eqn.~(\ref{eqn20d}) and/or Eqn.~(\ref{eqn12a}) may
be more difficult to exploit, as shall be discussed in
Section~\ref{subsec4.1}, but they can be used as a closure test for
the masses measured using the other constraints.

In typical SUSY decay chains we consider cases where the particles
$P$ and $Q$ can be either quarks/jets ($q$) or
leptons ($\ell$), leading to the following possible final-state
configurations: $\{P \equiv q,Q \equiv q\}$,
$\{P \equiv q,Q \equiv \ell\}$ and
$\{P \equiv \ell,Q \equiv \ell\}$. The second of
these configurations is particularly favourable from an experimental
point-of-view because in this case there is no ambiguity in assigning
particles to steps in the decay chains when constructing $\mct(P,P')$
and $\mct(Q,Q')$. In the case-studies presented below we shall
therefore focus on events with the final state $P \equiv q$
and $Q \equiv \ell$.

\subsection{Benchmarking on top events}\label{subsec4.1}

The mass measurement technique proposed above can be tested with
$t\bar{t}$ events in which both top quarks decay to leptons via
leptonically decaying $W$'s through the chain:
\begin{equation}
t \rightarrow bW \rightarrow b\ell\nu.
\label{eqn19top}
\end{equation}
Such events contain two symmetric two-step sequential two-body decay
chains, with invisible particles being produced at the end of each
chain. They therefore provide a suitable testbed for our mass
measurement technique, with $\delta \equiv t$, $\beta \equiv W$,
$\alpha \equiv \nu$, $P \equiv b$ and $Q \equiv \ell$. The main
notable difference between these events and SUSY events is that the
invisible particles are in reality approximately massless, however in
our analysis we shall not make this assumption. This approach has been
used previously to study alternative SUSY particle mass measurement
techniques \cite{Cho:2008cu,Burns:2008va}.

In order to evaluate the observables discussed in
Section~\ref{subsec4.0}, we generated with {\tt MC@NLO 3.3}
\cite{Frixione:2002ik,Frixione:2003ei} an inclusive sample of
$\sqrt{s}=14$ TeV LHC $t\bar{t}$ events with an input top mass of
172.5~GeV. The events were passed through the parameterised detector
simulation {\tt ACERDET} \cite{RichterWas:2002ch} which was modified to
reproduce the resolutions for leptons and jets given in
\cite{atlascsc}. For the tagging of $b$-jets, an efficiency of 60\%
was assumed, for a light jet rejection of 100. The total generated
sample was 2.2~M events, corresponding to an integrated luminosity of
approximately 3~fb$^{-1}$.

Events were selected with the following requirements:
\begin{enumerate}
\item $N_{\rm jet}$ $\geq$ 2, with $p_T(j_2)$ $>$ 40 GeV
\item $E_T^{miss}$ $>$ 30~GeV
\item $N_{\rm lep}$ $=$ 2, where ${\rm lep} = e/\mu$(isolated) and
$p_T(l_2)$ $>$ 20 GeV
\item At least two jets tagged as $b$ with $p_T>50$~GeV
\item Only one of the two possible sets of pairings of the two leptons
with the two leading $b$-jets should generate invariant mass values
which are both less than 175~GeV. This cut was intended to reduce the
experimental combinatorial background and was used only when
constructing observables which required the pairing of leptons and jets
from the same decay chain.
\end{enumerate}

Approximately 16100 (8300) events passed cuts 1-4 (1-5) respectively.
Of these 15200 (7400) were indeed events in which both hard $W$'s
decayed into a muon or electron. The remaining events contained at
least one tau lepton decaying leptonically into $e$ or $\mu$.

With the two $b\ell$ pairs, each corresponding to the decay of a
different top quark, one can construct the observables discussed in
Section~\ref{subsec4.0}. These observables are
$m(b^{(\prime)},\ell^{(\prime)})$, $\mct(b,b')$, $\mct(\ell,\ell')$
and $\mct([b\ell],[b'\ell'])$. Neglecting the mass of the $b$-quark,
the end-points in the distributions of these quantities lie at (from
Eqns.~(\ref{eqn20a})--(\ref{eqn20d})):
\begin{eqnarray}
m^{\rm max}(b,\ell)&=&\frac{\sqrt{[\mtsqr-\mwsqr]
[\mwsqr-\mnusqr]}}{\mw}=152.6~{\mathrm{GeV}},\label{eqn21a}\\
\mctmax(b,b')&=&\frac{\mtsqr-\mwsqr}{\mt}\equiv 135.0~{\mathrm{GeV}},\label{eqn21b}\\
\mctmax(\ell,\ell')&=&\frac{\mwsqr-\mnusqr}{\mw}\equiv 80.4~{\mathrm{GeV}},\label{eqn21c}\\
\mctmax([b\ell],[b'\ell'])&=&\frac{\mtsqr-\mwsqr}{\mt}
+\mt\left(\frac{\mwsqr-\mnusqr}{\mwsqr}\right)\equiv 307.5~{\mathrm{GeV}}.\label{eqn21d}
\end{eqnarray}
Accounting for $m(b)\neq0$ translates into shifts of less than $0.1\%$
in the end-point positions.

We show in Figure~\ref{fig:toppart} the distributions of the
observables at parton-level and at detector-level for events passing
the selection cuts in which both $W$'s decay into electron and
muons. All contransverse mass observables have been corrected for
transverse boosts according to the procedure discussed in
Section~\ref{subsec2.2}. It can be seen that the end-point structures
at parton-level are conserved at detector-level, modulo some
smearing. The enhancement observed at the lower limit of the $\mct(b,b')$
distribution is generated by the Jacobian peak at $\mct=\mctmin=2m(b)$
discussed in Section~\ref{sec2a} together with the $\mct=\mcy$
effect of the boost correction procedure discussed in the same
ssection. The $\mct(\ell,\ell')$ distribution in
Figure~\ref{fig:toppart} is relatively unaffected by the Jacobian
enhancement because $\mctmin=0$ for massless leptons. The dilepton
systems in these events receive large boosts from the $bb'$ recoil
however and so the boost correction procedure generates a prominent
$\mct=\mcy$ peak at $\mct=0$.

\FIGURE[ht]{
\epsfig{file=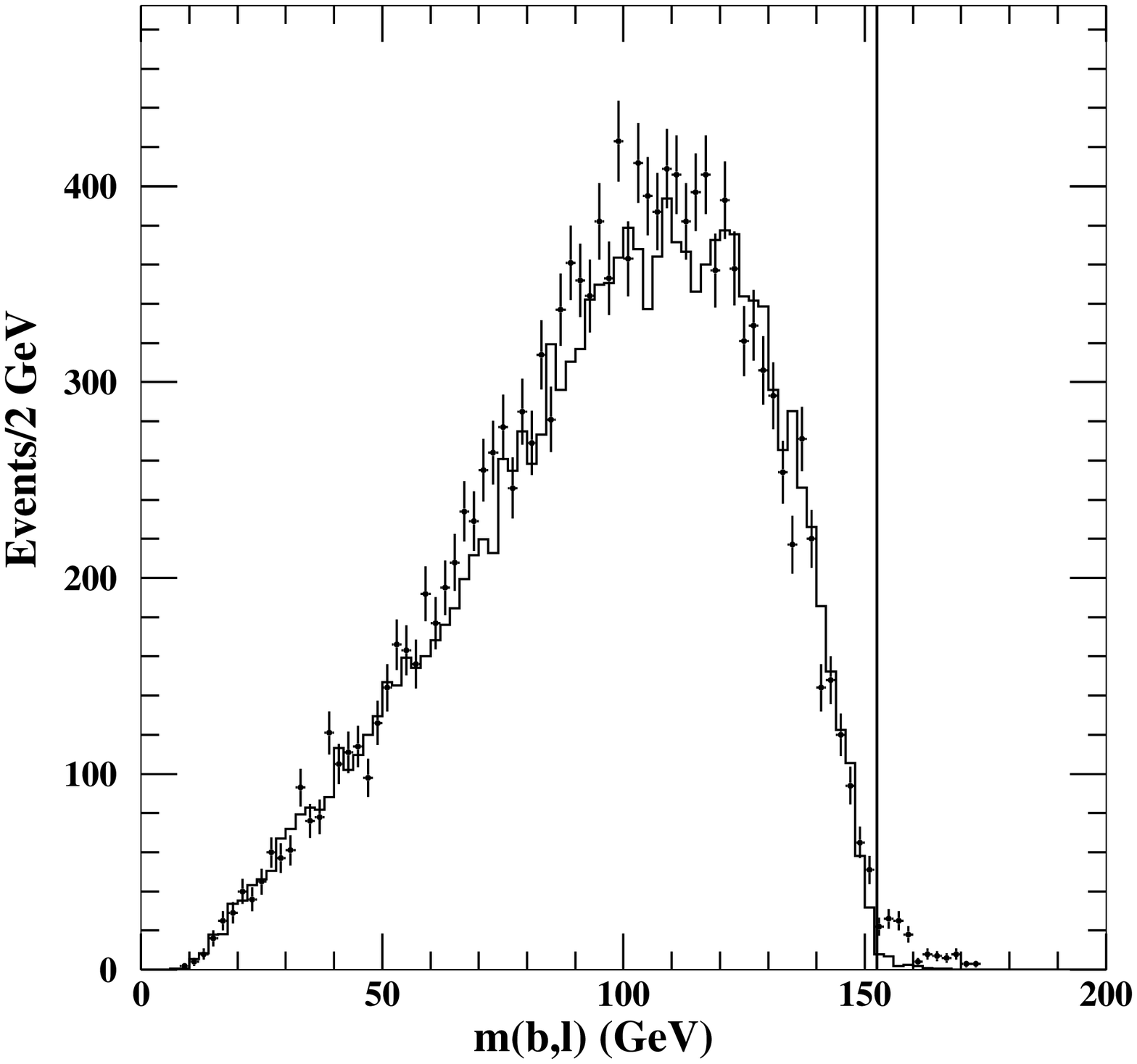,height=2.8in}
\epsfig{file=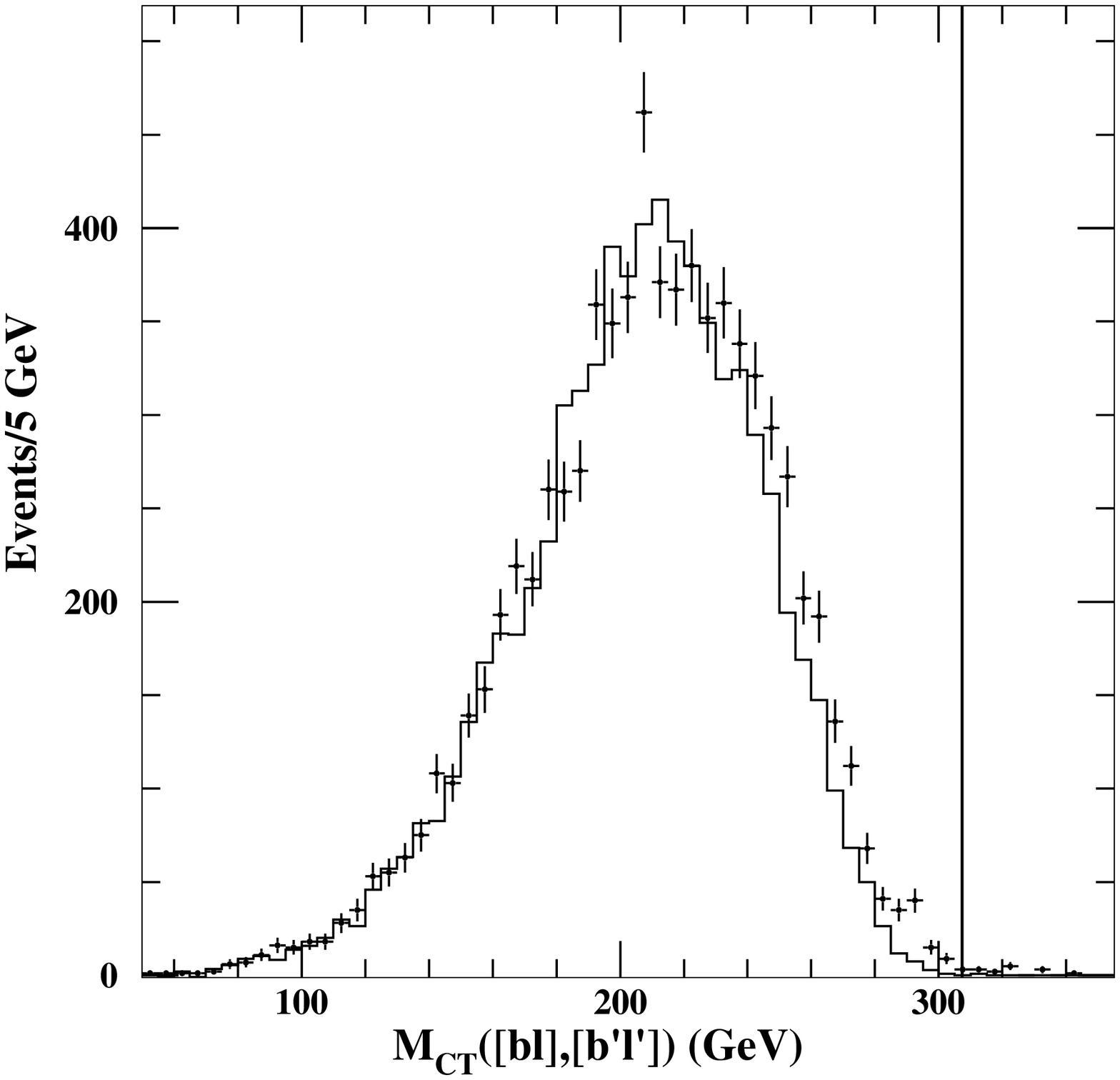,height=2.8in}
\epsfig{file=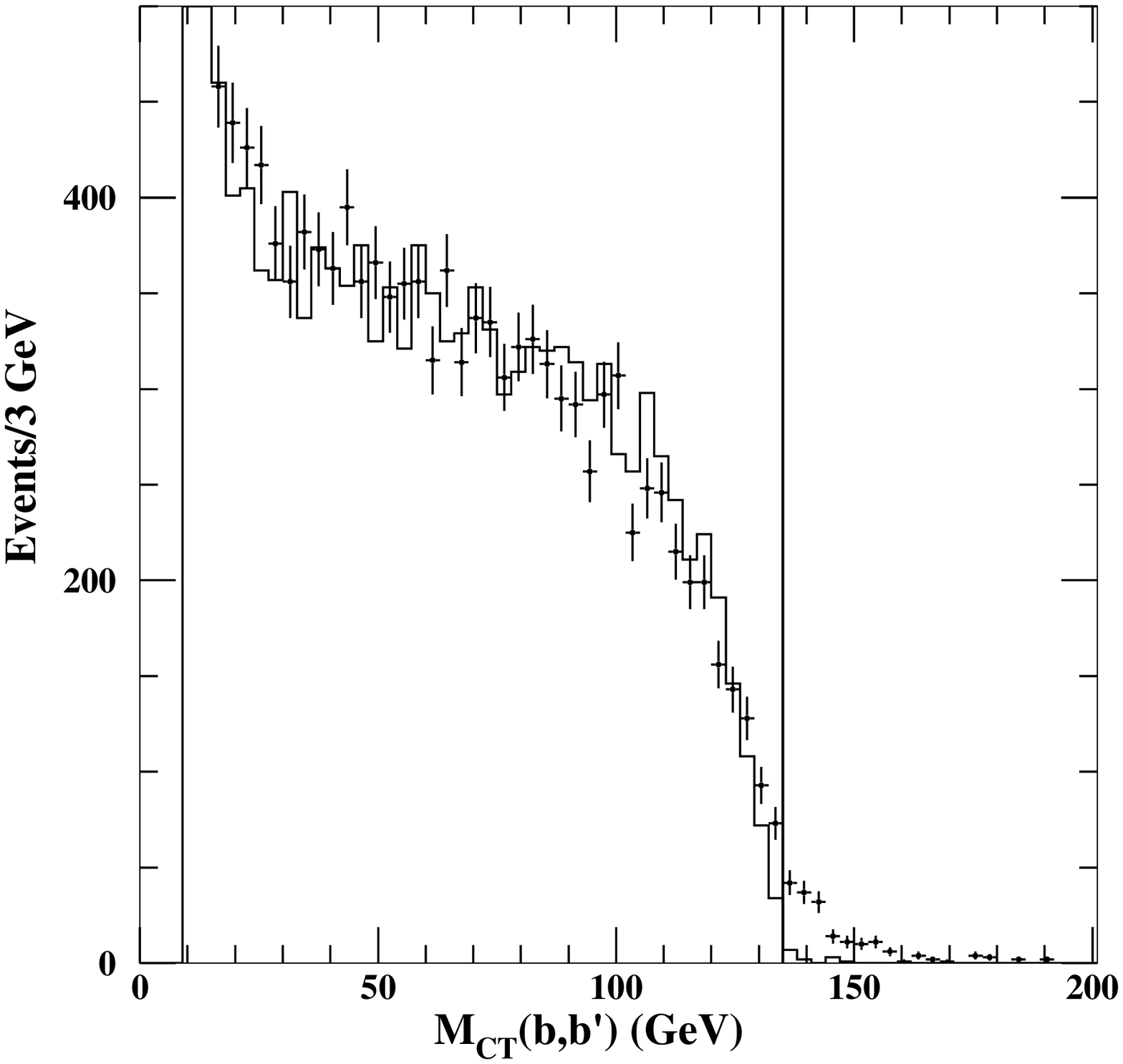,height=2.8in}
\epsfig{file=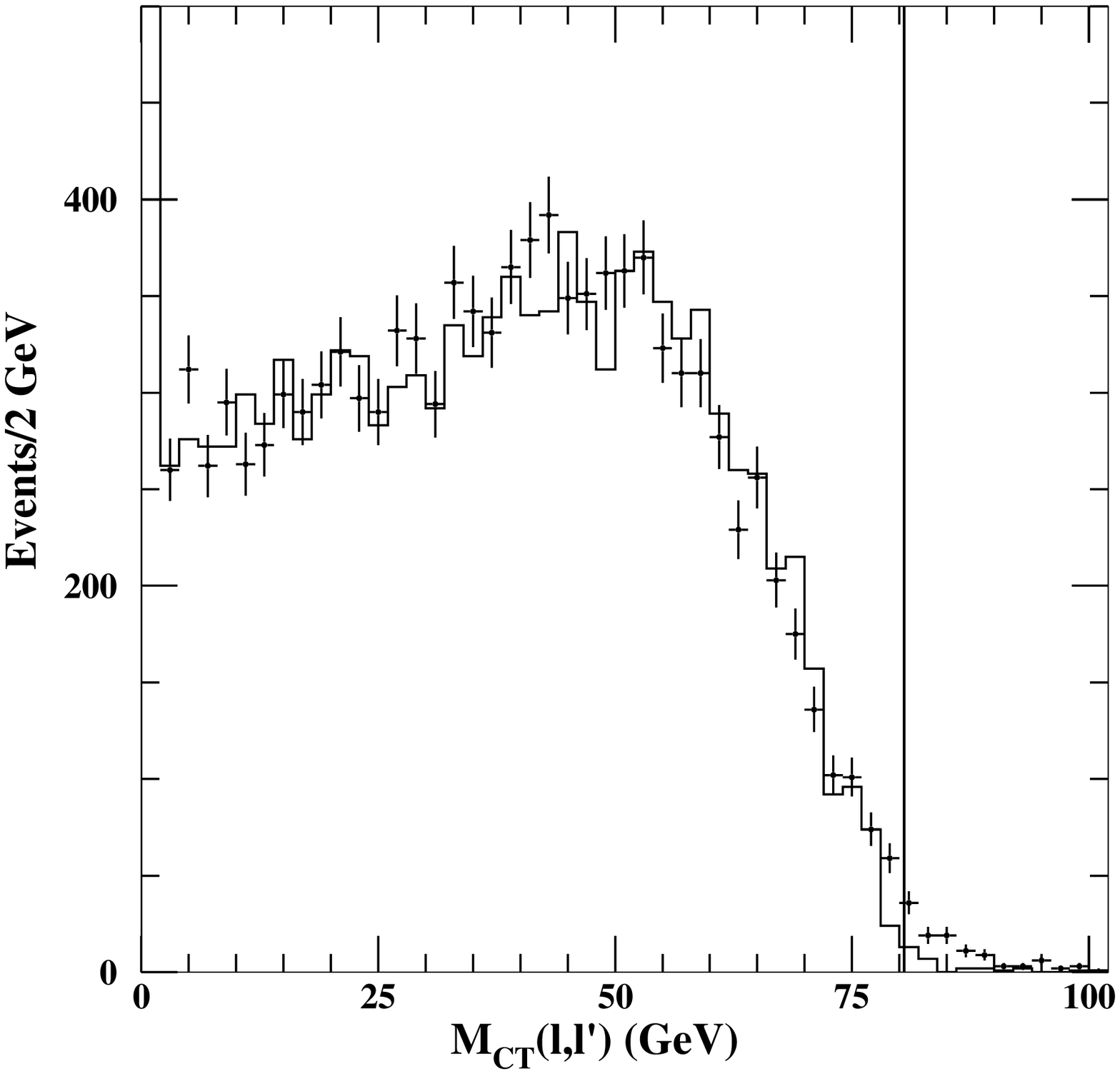,height=2.8in}
\caption{\label{fig:toppart} Distributions of
$m(b^{(\prime)},\ell^{(\prime)})$ (top-left),
$\mct([b\ell],[b'\ell'])$ (top-right), $\mct(b,b')$ (bottom-left) and
$\mct(\ell,\ell')$ (bottom-right) for $t\bar{t}$ events passing the
selection cuts where both leptons are generated directly from a W
decay. The histograms show the parton-level distributions while the
points with error-bars show the distributions after detector-level
smearing. The expected end-point positions are indicated with vertical
lines. The small populations of parton-level events lying beyond the
expected end-points arise from the natural width of the $W$.} }

\FIGURE[ht]{
\epsfig{file=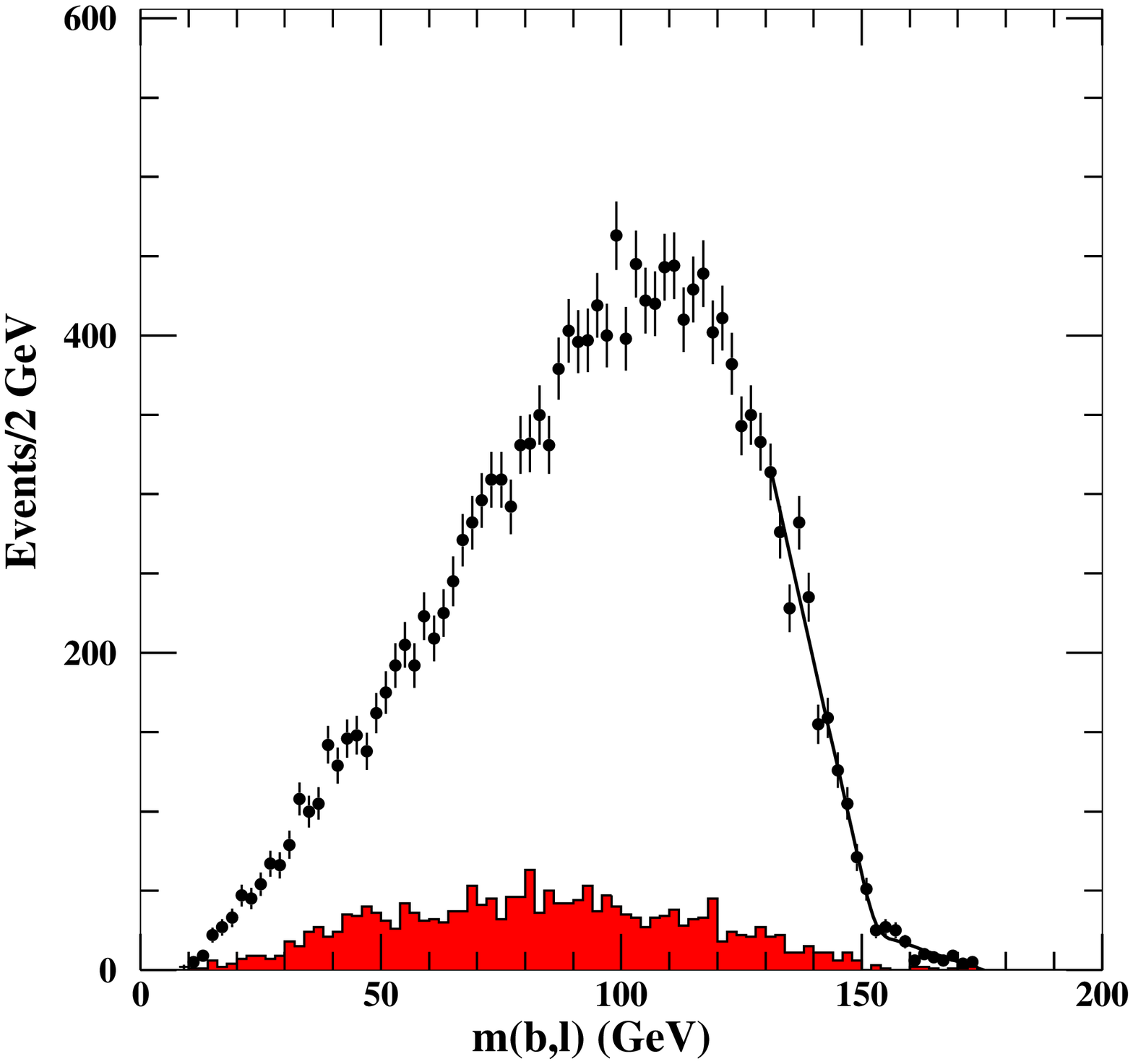,height=2.8in}
\epsfig{file=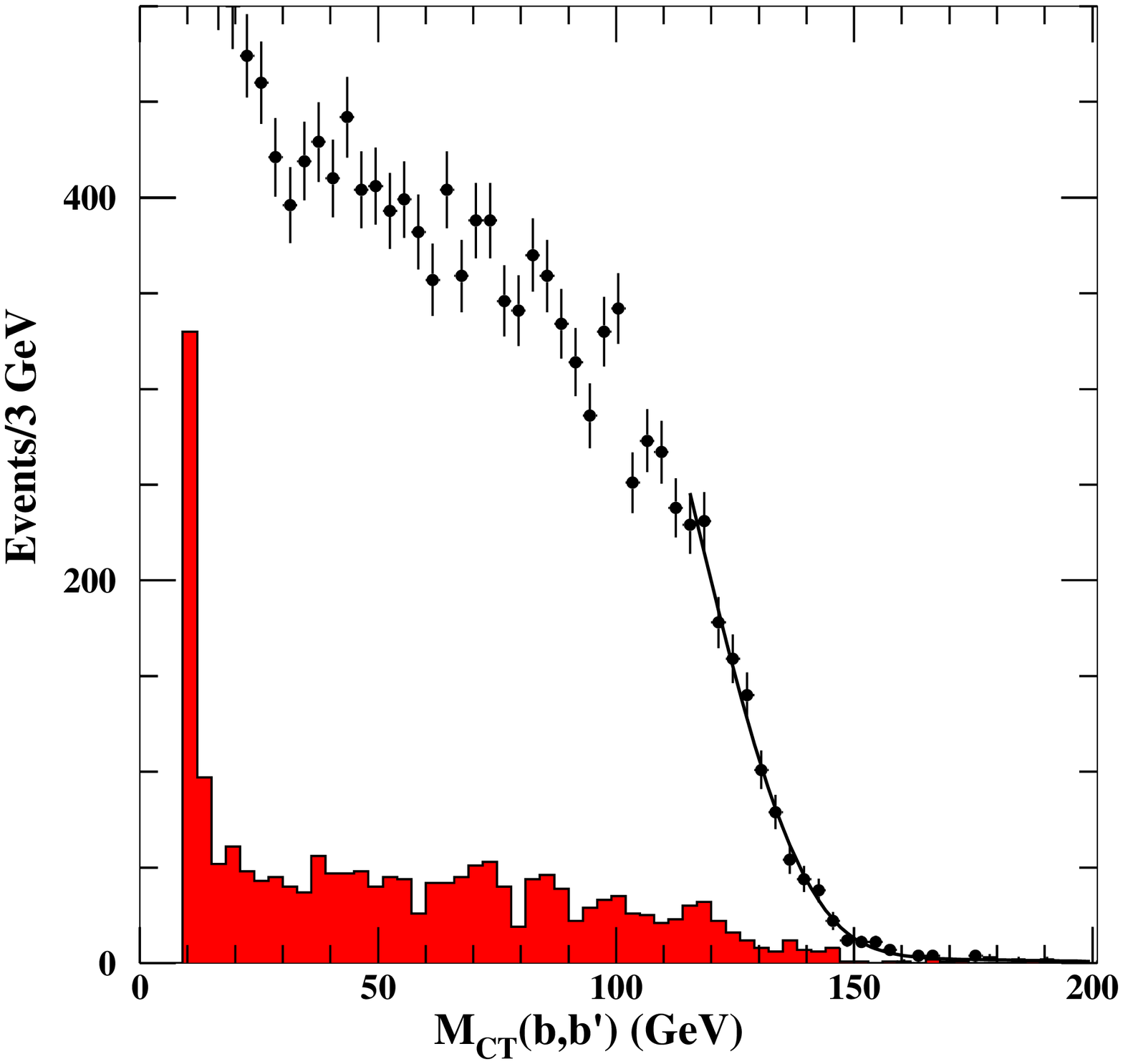,height=2.8in}
\caption{\label{figbb} Distributions of
$m(b^{(\prime)},\ell^{(\prime)})$ (left) and $\mct(b,b')$ (right) at
detector-level for $t\bar{t}$ events passing the selection cuts. The
dark (red) histogram indicates the distribution of events passing the
selection where one of the two leptons is not directly produced in the
decay of a $W$. The fit to the end-point function described in the
text is shown. } }

In order to explore in an approximate manner the potential precision
of mass measurements obtained with this technique, we fit the
end-points of the distributions with a linear function smeared by
detector resolution effects. Following Ref.~\cite{atlascsc} we use a
function $f(x)$ given by:
\begin{equation}
\label{eq:fitfunctionllq}
f(x) = \frac{1}{\sqrt{2\pi}\sigma} \int_0^{x^{\rm EP}} \exp\Big(-\frac{(x-x')^2}{2\sigma^2}\Big) \max\{A(x'-x^{\rm EP}),0\} {\rm\ d}x' + a+bx.
\end{equation}
Here $x$ is the observable under consideration, $x^{\rm EP}$
represents the end-point position, $\sigma$ represents the resolution of the
assumed gaussian detector smearing, $A$ is the slope of the
distribution before smearing, and $a$ and $b$ are parameters
describing an assumed linear background distribution. The latter
distribution helps to take into account the effects of both
combinatorial background from incorrect assignment of visible
particles to decay chains and non-gaussian tails in the experimental
resolution.

\FIGURE[ht]{
\epsfig{file=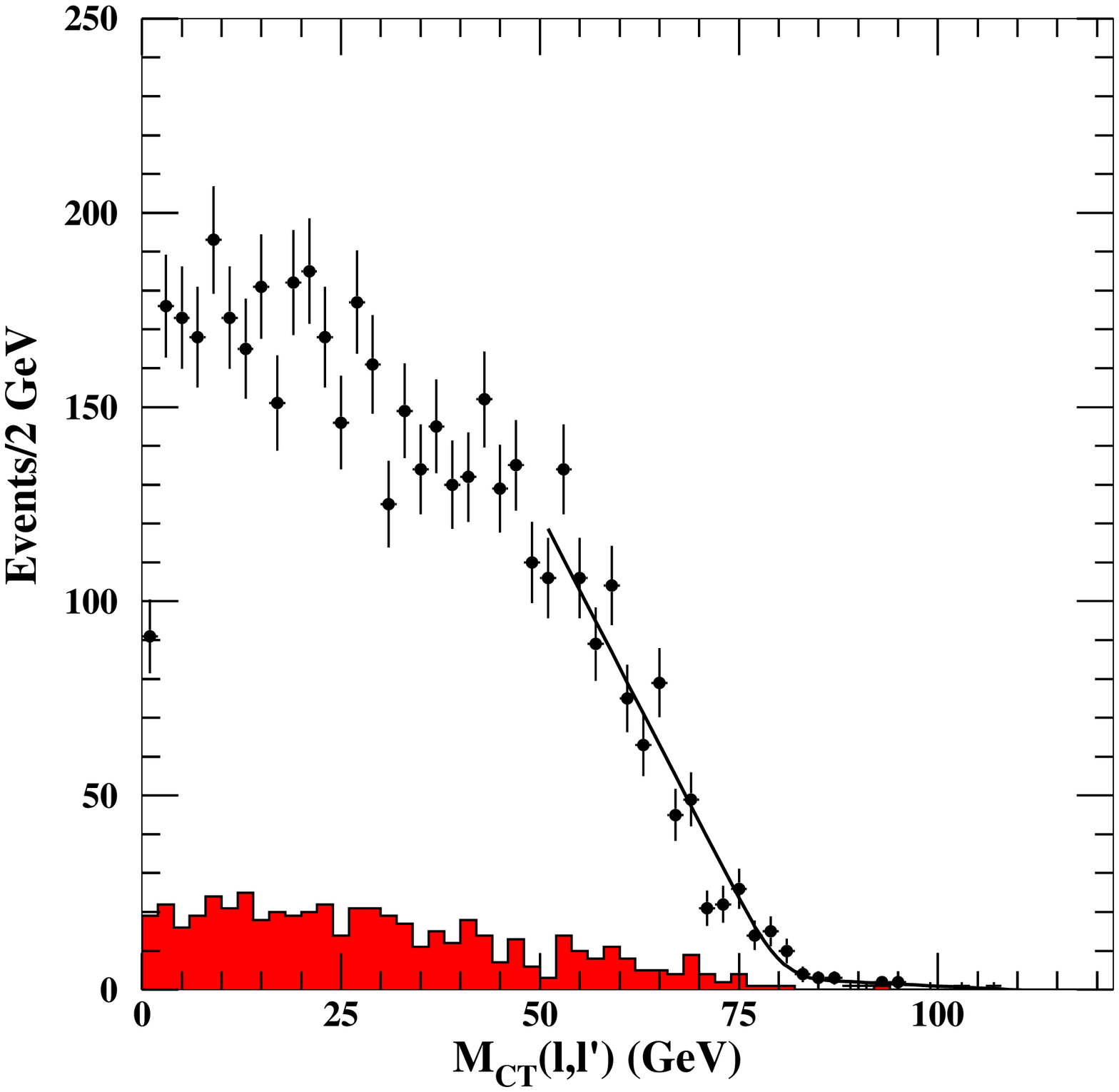,height=2.8in}
\epsfig{file=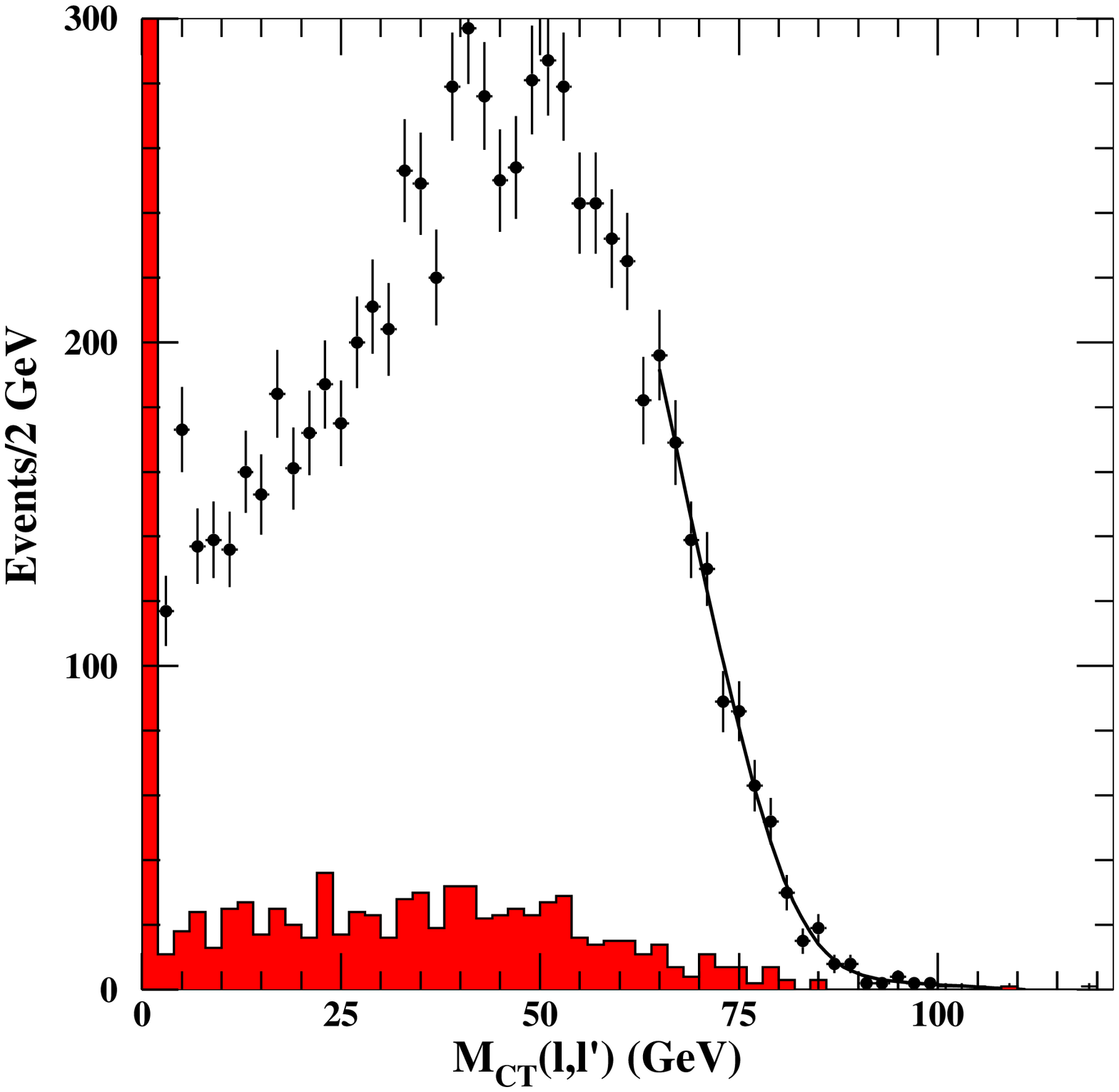,height=2.8in}
\caption{\label{figll} Distributions of $\mct(\ell,\ell')$ for $\axl>0$
(left), and $\axl<0$ (right) at detector-level for $t\bar{t}$ events
passing the selection cuts. The dark (red) histogram indicates the
distribution of events passing the selection where one of the two
leptons is not directly produced in the decay of a $W$. The fit to the
end-point function described in the text is shown.} }

For the observables $m(b^{(\prime)},\ell^{(\prime)})$ and
$\mct(b,b')$, the fits to the distributions for all detector-level
events passing the cuts are shown in Figure~\ref{figbb}, with the
irreducible background of $t\bar{t}$ events where at least one of the
leptons is not directly produced in a $W$ decay shown in grey
(red). The fit function reproduces well the observed shape, and the
value of the resolution parameter $\sigma$ obtained from the fit is in
good agreement with the actual value of the smearing used in the
detector simulation.

The situation is somewhat more complex for the $\mct(\ell,\ell')$
observable. In this case one has two populations. If $\axl>0$ only a
very small transverse boost correction is applied to $\mct$, using
$\eog=E_{\rm cm}$. Therefore the experimental end-point resolution is
to a good approximation just the resolution of the lepton $p_T$
measurement (of order 1~GeV), plus the end-point smearing arising from
the $W$ natural width (of order 2-3 GeV). If $\axl<0$ however, $\mct$
is corrected using the $p_T$ of the hadronic recoil, resulting in a
significantly larger resolution of order 9-10~GeV. The two
configurations must therefore be fitted separately. The distributions
are shown in Figure~\ref{figll}, for $\axl>0$ ($\axl<0$) on the left
(right). From the figure one can also observe that the gradient of the
$\axl>0$ distribution near the end-point is smaller, and it was
necessary in this case to fix the experimental resolution to 3~GeV in
order to obtain an acceptable fit.

\FIGURE[b]{
\epsfig{file=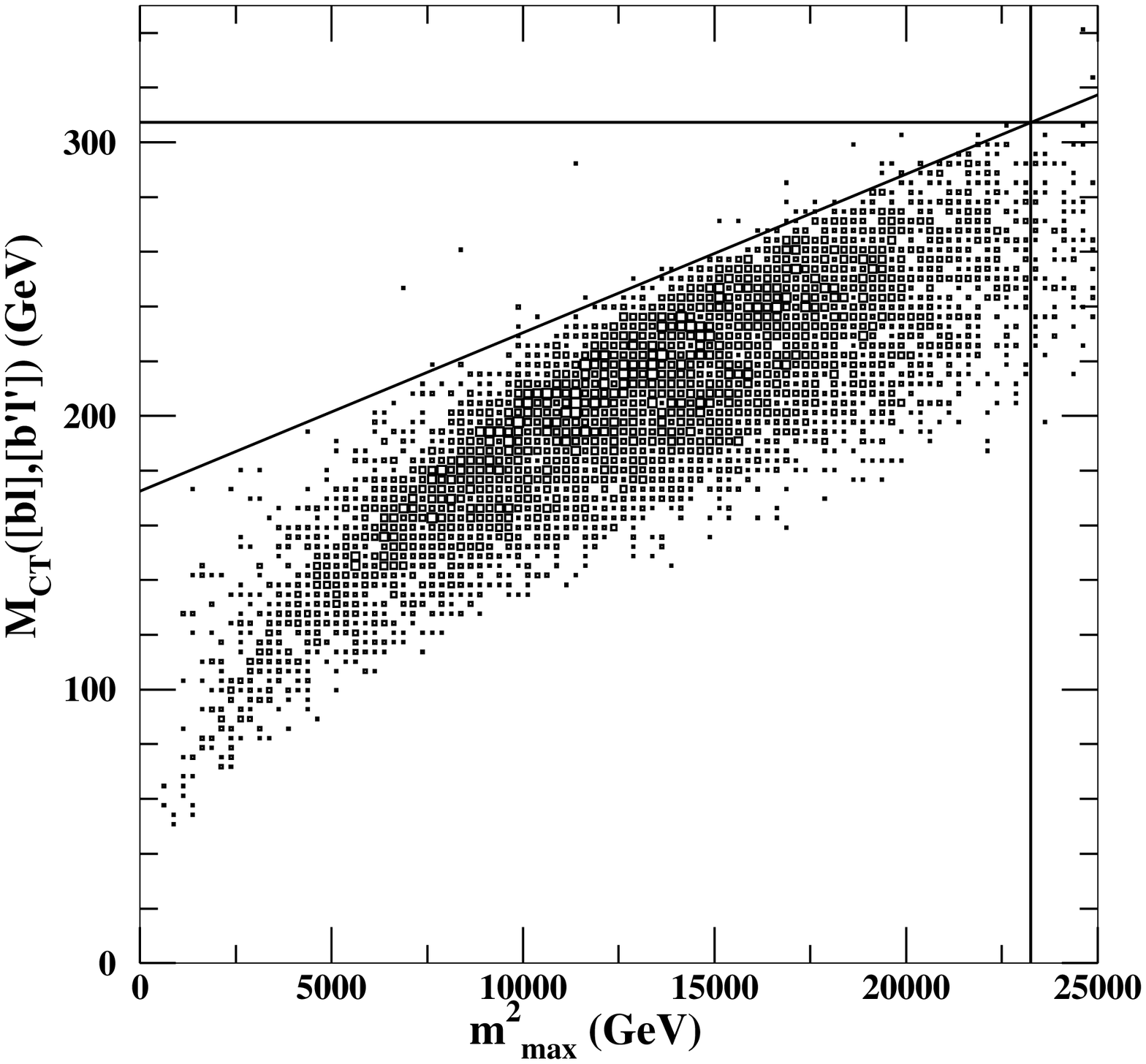,height=4in}
\caption{\label{fig4} Two-dimensional distribution in the
$\mct([b\ell],[b'\ell'])$ versus $\mmax^2$ plane of detector-level
$t\bar{t}$ events passing the selection cuts. The extremal values of
the two observables, given by Eqns.~(\ref{eqn21d}) and~(\ref{eqn21a})
are denoted by horizontal and vertical lines respectively. The
dependence of $\mctmax([b\ell],[b'\ell'])$ on $\mmax^2$ given by
Eqn.~(\ref{eqn18}) is denoted by the diagonal line.} }

Measurement of the end-point in the $\mct([b\ell],[b'\ell'])$
distribution presents further challenges due to the concave shape of
the distribution near to the end-point, seen in
Figure~\ref{fig:toppart}(top-right). For this end-point the assumption
of a linear shape breaks down, primarily due to the depopulation of
the $\mct([b\ell],[b'\ell'])$ versus $\mmax^2$ plane near
$\mmax^2=(m^{\rm max}(b,\ell))^2$ seen in Figure~\ref{fig4}. An
alternative strategy for constraining the masses with
$\mct([b\ell],[b'\ell'])$ would be to measure the dependence of
$\mctmax([b\ell],[b'\ell'])$ on $\mmax^2$, as discussed in
Section~\ref{subsec3.1}. This could be accomplished in practice by
constructing $\mct([b\ell],[b'\ell'])$ histograms of those events
which pass a cut on $\mmax^2$. Unfortunately however this procedure is
also complicated by the presence of concave end-points, as can be seen
in Figure~\ref{figproj}. Further progress with this specific element
of the contransverse mass technique will likely require a dedicated
study of end-point shapes, which is outside of the scope of this
paper.  Because of these considerations we will not use the
measurements of $\mctmax([b\ell],[b'\ell'])$ in the following mass
measurement study. Nevertheless such constraints could be useful for
validating mass measurements obtained from the other
observables. Similar considerations apply when attempting to
use the non-boost-corrected $\mct(b,b')$ and $\mct(\ell,\ell')$
observables to measure the dependence of $\mctmax(b,b')$ and
$\mctmax(\ell,\ell')$ on $p_b$, as shown in Figure~\ref{fig-mctpb},
and this technique will also not be considered further here.

\FIGURE[ht]{
\epsfig{file=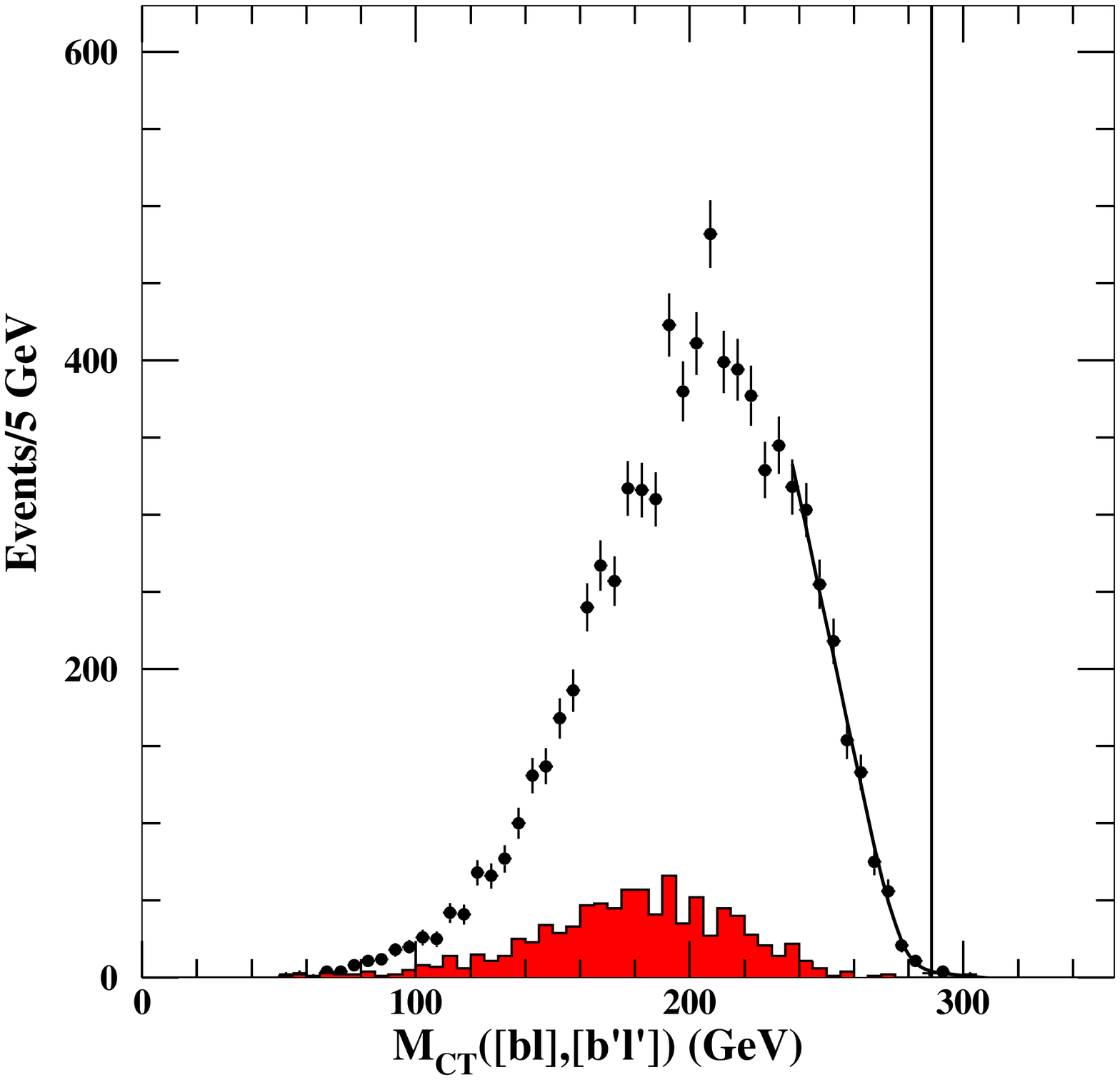,height=2.8in}
\epsfig{file=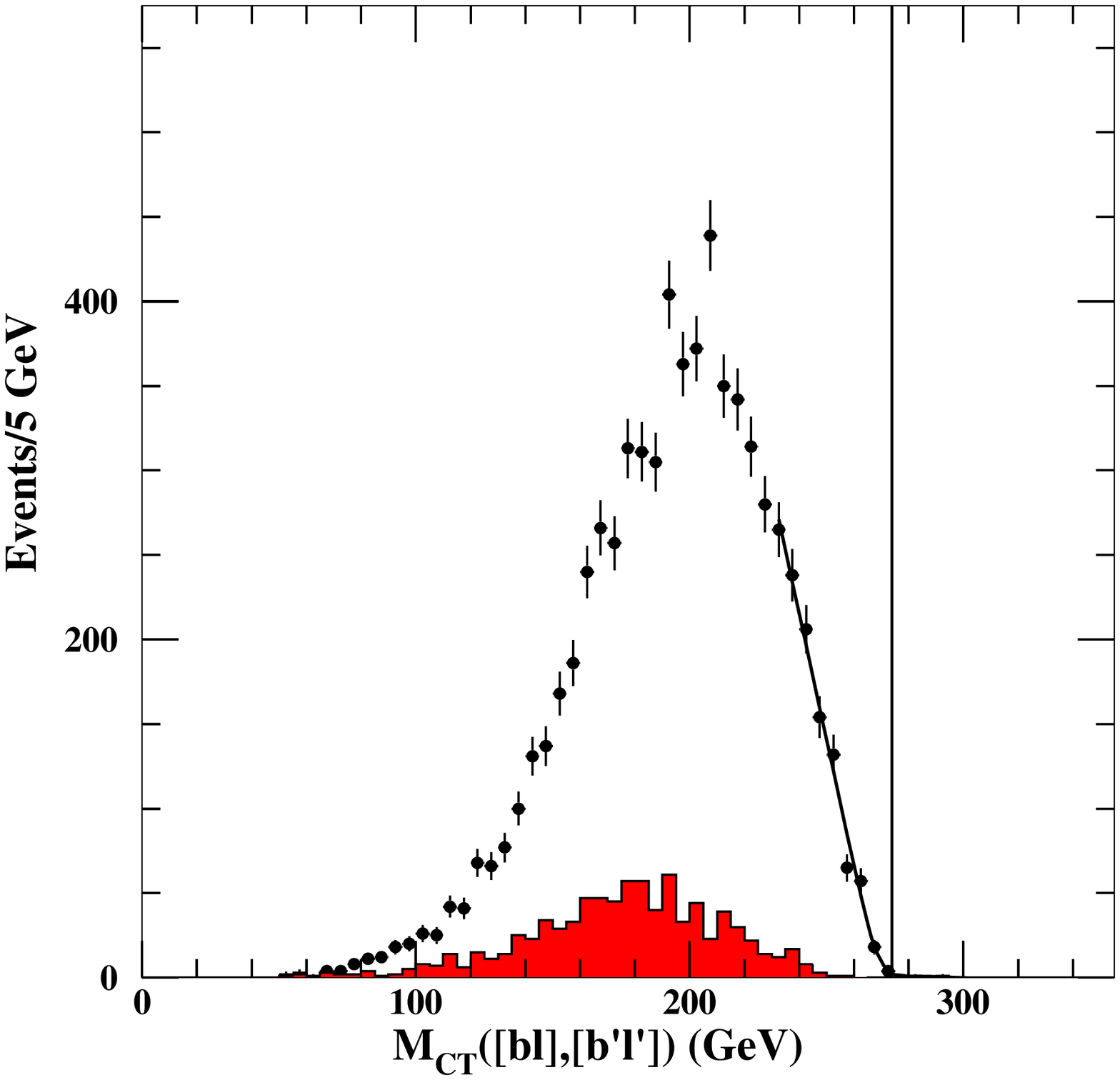,height=2.8in}
\caption{\label{figproj} Distributions of $\mct([b\ell],[b'\ell'])$
for detector-level $t\bar{t}$ events passing the selection cuts after
requiring additionally that $\mmax^2<20000$~GeV$^2$ (left) and
$\mmax^2<17500$~GeV$^2$ (right). The vertical lines indicate the
expected end-point positions.  } }

\FIGURE[ht]{
\epsfig{file=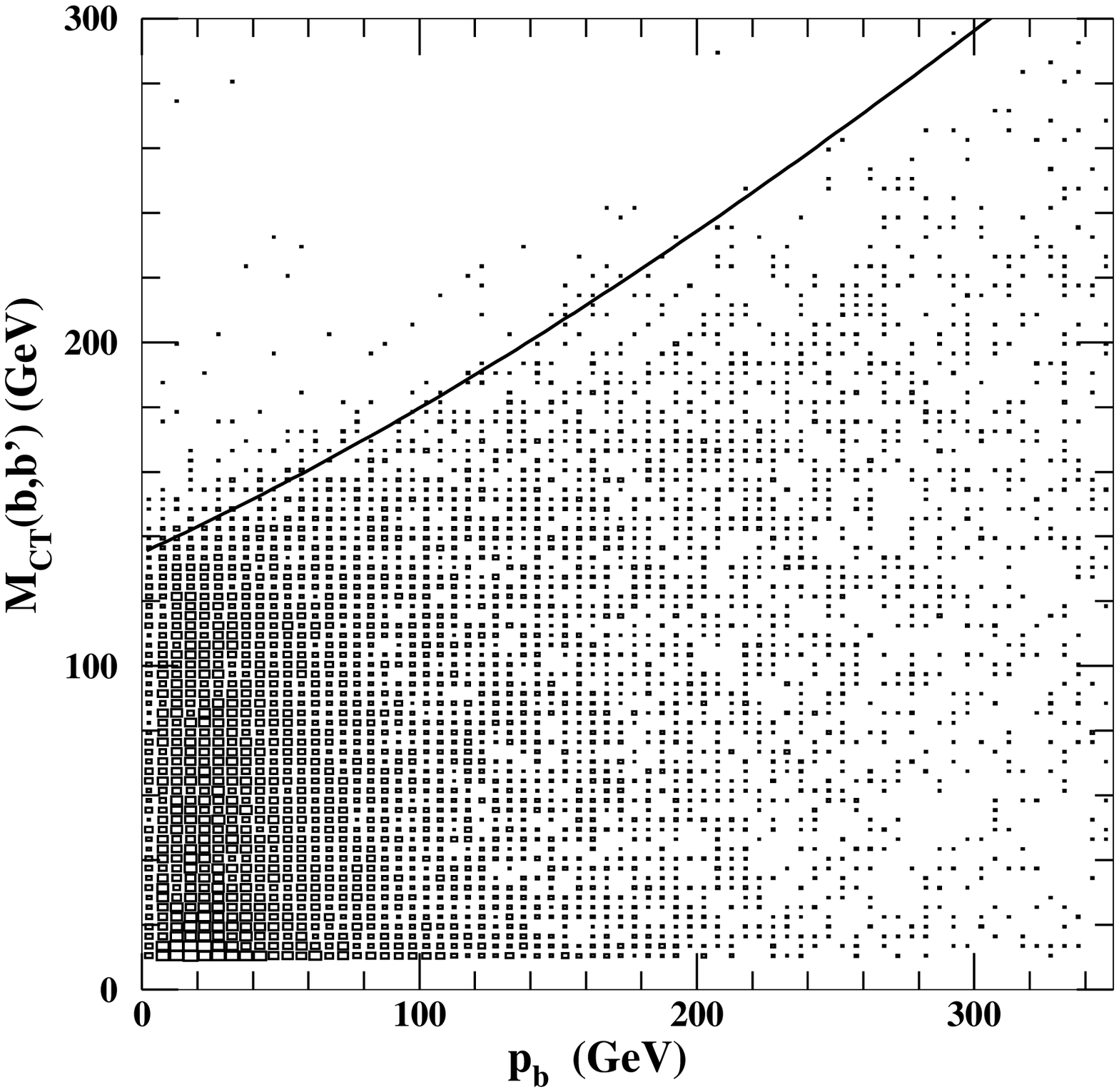,height=2.8in}
\epsfig{file=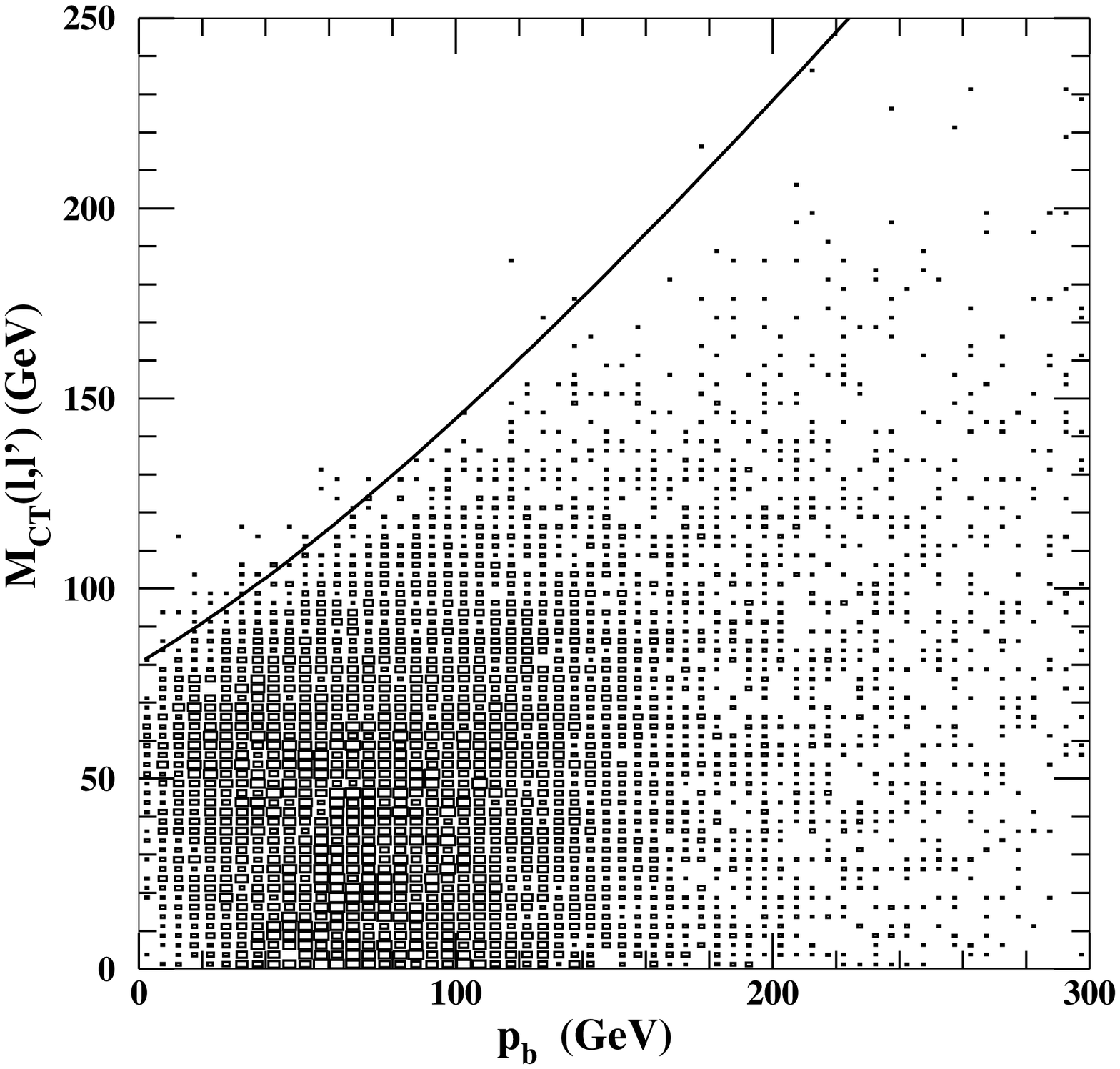,height=2.8in}
\caption{\label{fig-mctpb} Two-dimensional distributions in the
$\mct(b,b')$ versus $p_b$ plane (left) and $\mct(\ell,\ell')$ versus
$p_b$ plane (right) of detector-level $t\bar{t}$ events passing the
selection cuts. The dependence of $\mctmax$ on $p_b$ given by
Eqn.~(\ref{eqn12d}) is denoted by the curved line in each case. The
small population of events lying beyond $\mctmax$ in the left-hand
figure arises from the finite detector-level b-jet energy resolution,
which degrades at lower energy.} }

The results of the end-point fits are listed in Table~\ref{tabfit},
where the first uncertainty is the statistical uncertainty from the
{\tt MINUIT} \cite{James:1975dr} fitting program for the chosen fit
interval, the second is the systematic uncertainty obtained by varying
the fit interval and the third uncertainty is the correlated
systematic uncertainty derived from assumed energy scale uncertainties
of 1\% for $b$-jets and 0.1\% for leptons \cite{atltdr}. The quoted
uncertainties should be considered approximate and could be improved
with the use of better end-point fitting functions, for instance
templates derived from Monte Carlo simulation studies.

\TABLE[ht]{
\caption{End-point positions in GeV. The first uncertainty is
statistical, while the second and third are respectively the
uncorrelated systematic and correlated energy scale uncertainties. The
expected end-point positions from Eqns.~(\ref{eqn21a}),~(\ref{eqn21b})
and~(\ref{eqn21c}) are listed in the column labelled `Truth'. The
assumed integrated luminosity is 3~fb$^{-1}$.}
\bigskip
\label{tabfit}
   \begin{tabular}{l|c|c}
   End-point &  Truth (GeV) & Measured  (GeV) \\
   \hline
$m^{\rm max}(b,\ell)$   & 152.6  &  $152.8 \pm 1.7 \pm 1 \pm 0.8$ \\
$\mctmax(b,b')$ &   135.0 &  $137.7 \pm 3.6 \pm 3 \pm 1.4$ \\ 
$\mctmax(\ell,\ell')$ $(\axl<0)$  & 80.4  & $80.2 \pm 0.5 \pm 1 \pm 0.1$ \\
$\mctmax(\ell,\ell')$ $(\axl>0)$  & 80.4  & $81.2 \pm 1.7 \pm 2 \pm 0.4 $ \\
   \end{tabular}
}

Based on the end-point measurement uncertainties listed in
Table~\ref{tabfit} it is possible to evaluate the achievable
precisions for measuring the masses of the top quark, $W$ and
neutrino. We use the technique described in Ref.~\cite{Nojiri:2005ph},
where for each end-point measurement we generate a set of
pseudo-experiments by sampling from a gaussian distribution centred on
the nominal value of the end-point position of width equal to the estimated
measurement precision. We assume that the measurements are
uncorrelated with the exception of the energy scale uncertainties,
which are assumed to be fully correlated. For each pseudo-experiment
we calculate the value of $\mt$ according to Eqn.~(\ref{eq:ma})
and hence calculate $\mw$ and $\mnu$ from Eqns.~(\ref{eqn21b})
and~(\ref{eqn21c}).  

\FIGURE[ht]{
\epsfig{file=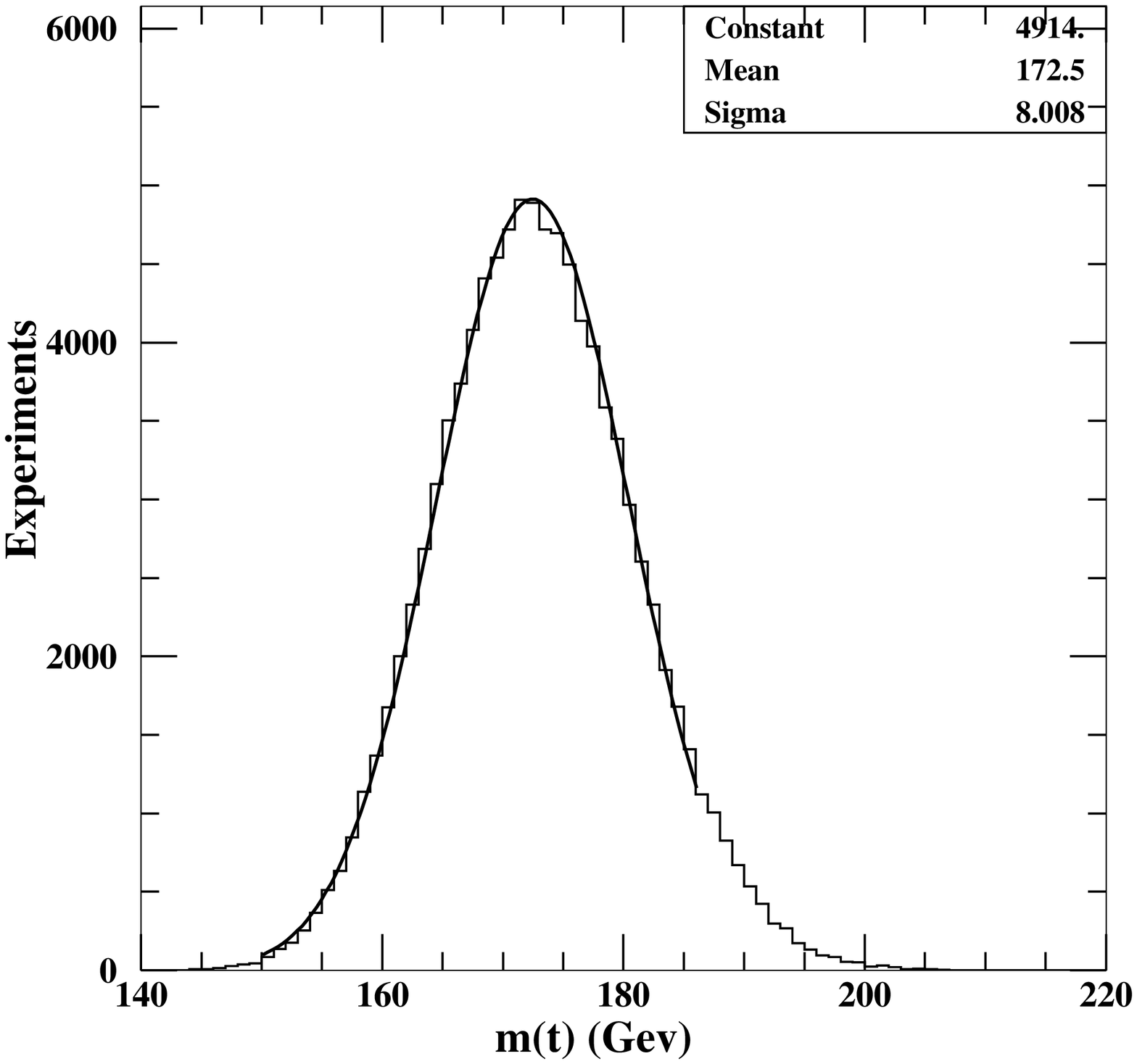,height=2.8in}
\epsfig{file=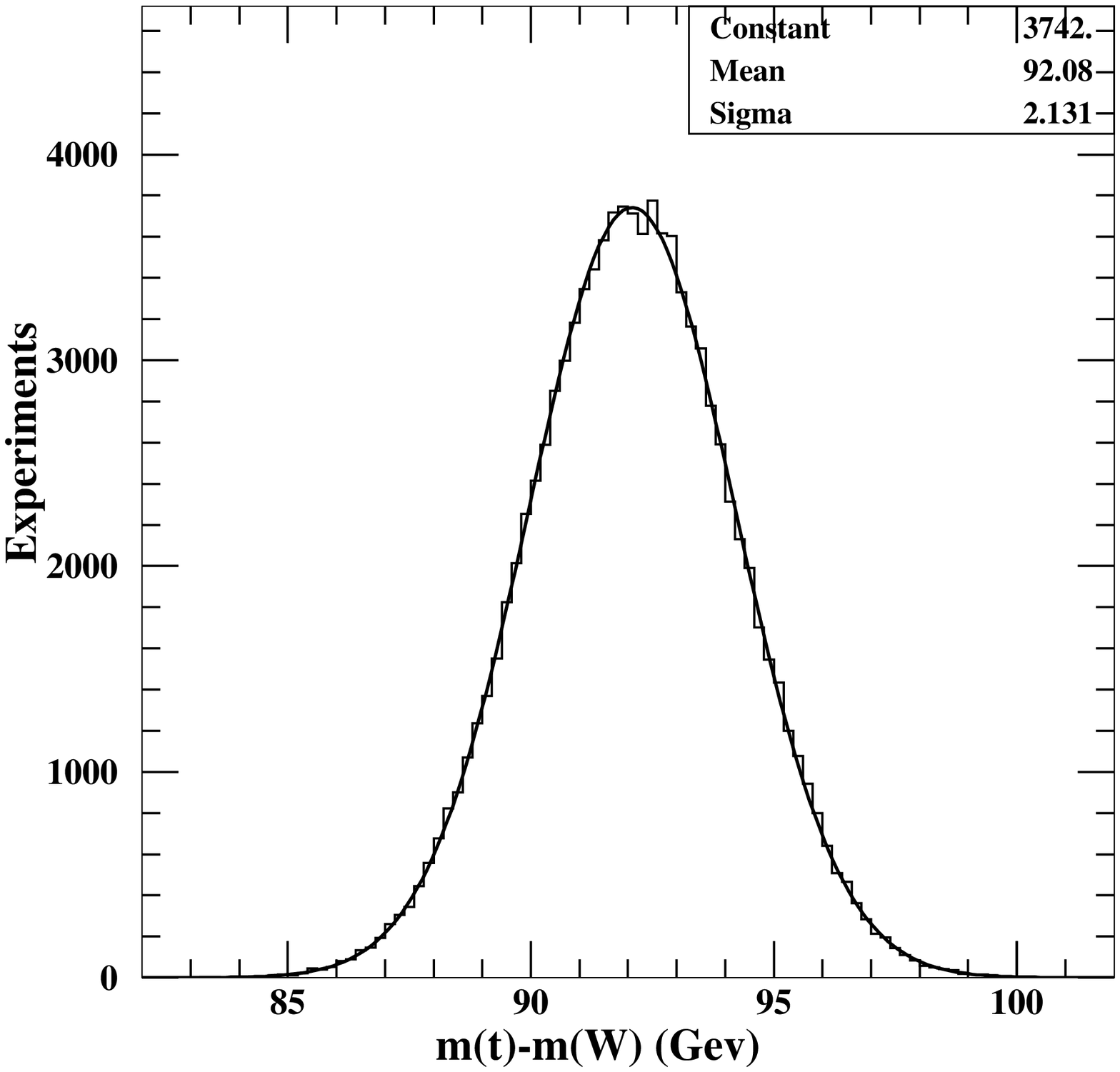,height=2.8in}
\caption{\label{fig:massf} Distributions of the calculated top mass
(left) and $\mt-\mw$ mass difference for 100k experiments. The assumed
statistics is 3~fb$^{-1}$.  } }

The distributions of the measured $\mt$ values and $\mt-\mw$ mass
differences are shown in Figure~\ref{fig:massf} for a set of 100000
pseudo-experiments. The precision of the top quark mass measurement is
$\sim$8~GeV, while the uncertainty on the measurement of $\mt-\mw$ is
$\sim$2~GeV.  A 95\%(68\%) upper limit on the neutrino mass of 30(16)
GeV is obtained. These results may appear to be disappointing when
compared with the $\sim$1~GeV $\mt$ precision expected to be obtained
at the LHC from semileptonic $t\bar{t}$ events for the same
assumptions on $b$-jet energy scale uncertainty \cite{atlascsc}. Here
however we have made no assumptions about the masses of the $W$ or
neutrino and so 8 GeV is the stand-alone precision with can be
obtained with the technique. The end-point measurements used in this
technique are primarily sensitive to mass differences and so if the
mass of the $W$ were assumed to be known the precision of the
measurement of $\mt$ would improve to $\sim$2~GeV, dominated by the
systematics associated with the very crude end-point fitting function
used for this study.

%% file: squark.tex
\subsection{A SUSY example}\label{subsec4.2}

Having demonstrated the proposed mass measurement technique with
$t\bar{t}$ events let us now apply the same technique to a SUSY model
generating events with a similar final state. 
An example of such a
SUSY model is an MSSM model with a left-handed slepton doublet lighter
than the chargino. In this case events with the decay chain
\begin{equation}
\tq\rightarrow q\tchi^\pm_1\rightarrow q \ell \tilde{\nu} \rightarrow
q \ell \nu \lsp
\label{decnu}
\end{equation}
appearing in both legs of the event can be produced. The invisible
sparticle at the end of the chain is in this case the sneutrino, since
both of its decay products are undetected. This decay chain maps onto
Eqn.~(\ref{eqn19}) with $\delta \equiv \tq$, $\beta \equiv \tchi^\pm_1$,
$\alpha \equiv \tilde{\nu}$, $P \equiv q$ and $Q \equiv \ell$.

The decay (\ref{decnu}) is however not the only decay yielding the
final state of interest, with a quark, a lepton and one or more
invisible particles on each leg. We consider the following decay
chains:
\begin{eqnarray}
\tq\rightarrow q\tchi^\pm_1 &\rightarrow& q\nu \tilde{\ell}\rightarrow q\nu \ell \lsp \label{decl} \\
\tq\rightarrow q\tchi^\pm_1 &\rightarrow&  q\lsp W \rightarrow q\lsp \ell \nu
\label{dec3} \\
\tq\rightarrow q\tchi^\pm_1 &\rightarrow&  q\lsp W \rightarrow q\lsp \ell \nu
\label{dec4}
\end{eqnarray}

For all of these decays, formulas (\ref{eqn20a})--(\ref{eqn20d}) are
valid, provided that $m(\alpha)$ in (\ref{eqn20a}), (\ref{eqn20c}) and
(\ref{eqn20d}) is replaced by $m_{\rm min}(\alpha)$, defined as the
minimum mass of the ``pseudo-particle'' composed of all of the
invisible particles in the decay. Analytical expressions for $m_{\rm
min}(\alpha)$ in terms of the masses of the particles involved in the
decays are given in Appendix B. These can be used in
Eqns.~(\ref{eqn20a}) and~(\ref{eqn20d}).

The case of Eqn.~(\ref{eqn20c}) deserves special comment. In this
case for decay chains (\ref{decl}), (\ref{dec3}) and (\ref{dec4})
one invisible particle is upstream of the lepton and one
downstream. It is therefore not possible to correct for the upstream
momentum, since it is not possible to separate the momentum of the
neutrino and of the $\lsp$. However, if one performs the boost
correction assuming that all of the observed missing transverse
momentum is downstream of the lepton, the distributions for
$\mctmax(Q,Q')$ still possess end-points at positions given by
Eqn.~(\ref{eqn20c}) with the $m_{\rm min}(\alpha)$ values defined in
Appendix B.

Based on measurements of the end-point positions $k_1$, $k_2$ and
$k_3$, one can calculate the masses of the squark and of the chargino
independently from the decay mode of the $\tchi^\pm_1$.  This is a
remarkable achievement, as it shows that it is possible to perform an
absolute measurement of the chargino mass through its leptonic decay
notwithstanding the fact that two invisible sparticles are present in
the decay of each chargino. The interpretation of the meaning of the
measured $m(\alpha)$ {\it is} dependent on the chargino decay mode,
and the analysis based on the measurement of $k_1$, $k_2$, $k_3$ and
$k_4$ does not allow the discrimination of the different expressions
given in Appendix B.

It should be noted that the mass hierarchy and coupling structure
implied by the presence of the decay chain given by Eqn.~(\ref{decnu})
imply also the existence, with a significant branching ratio, of the
chain:
\begin{equation}
\tq\rightarrow q\tchi^0_2\rightarrow q \ell \tl_L \rightarrow q \ell \ell \lsp,
\label{decll}
\end{equation}
which can be easily selected by requiring the presence of two leptons
with the same flavour and opposite sign. This `golden channel'
for SUSY mass measurement at the LHC can potentially provide
additional information regarding the masses of sparticles involved in
the chargino decay chain. In the following we shall not show a
complete analysis along these lines, which has been already developed
in detail in e.g. Refs.~\cite{Bachacou:2000zb,
Allanach:2000kt,Weiglein:2004hn}. We shall limit ourselves instead to
showing for our example model that the invariant mass of two OS-SF
leptons does give a characteristic end-point structure from the
\mbox{$\tchi^0_2\rightarrow\ell\tilde{\ell}$} decay.
Starting from this end-point, and combining it with the hard jets in
the events it is possible to measure the masses of $\tq$, $\tchi^0_2$
$\tilde{\ell}$ and $\lsp$
\cite{Bachacou:2000zb,Allanach:2000kt,Weiglein:2004hn}.  It is then
straightforward to insert the measured values of these masses into
Eqns.~(\ref{eqn20a})--(\ref{eqn20d}) and observe that under the
assumption of decays (\ref{decl}), (\ref{dec3}), and (\ref{dec4}) the
mass measurements from the two analyses are inconsistent.

In order to explore the feasibility of this measurement technique, we
used {\tt HERWIG 6.5} \cite{Corcella:2000bw, Moretti:2002eu} to
generate events from a toy MSSM model incorporating the mass hierarchy
present in decay chain (\ref{decnu}). The masses of all of the squarks
were set to 500~GeV and those of all the sleptons to 150~GeV. The
three gaugino masses $M_1$, $M_2$ and $M_3$ were set respectively to
120, 250 and 520~GeV while the higgsino mass parameter $\mu$ was set
to 400~GeV, $\tan\beta$ to 10, and $m_A$ to 400~GeV. The trilinear
couplings were set to zero. The relevant sparticle masses, as
calculated by {\tt ISASUSY 7.75} \cite{Paige:2003mg} are listed in
Table~\ref{tabsusy}. The expected end-point positions for the chains
(\ref{decnu}) and~(\ref{decl}) are listed in Table~\ref{tab:edges},
using the results of the discussion above regarding the treatment of
$m(\alpha)$ in decay chain~(\ref{decl}).
\TABLE[ht]{
\caption{Masses of the relevant sparticles for the example MSSM point.}
\bigskip
\label{tabsusy}
   \begin{tabular}{l|c|l|c}
 Parameter    & Value (GeV)   & Parameter & Value (GeV) \\
   \hline
$m(\tg)$           &  520.0 &  $m(\tilde{u_L})$       &  503.4   \\
$m(\tilde{e_L})$   &  157.1 &  $m(\tilde{\nu_e})$     &  135.7   \\
$m(\tchi^\pm_1)$   &  231.5 &  $m(\tchi^0_2)$  &  232.0   \\
$m(\tchi^0_1)$     &  117.2 &   & \\
   \end{tabular}
}
\TABLE[ht]{
\caption{Expected end-point positions in GeV for the decay chains
(\ref{decnu}) and (\ref{decl}).}
\bigskip
\label{tab:edges}
   \begin{tabular}{l|c|c}
 End-point    & Position: Chain (\ref{decnu})  & Position: Chain (\ref{decl})\\
   \hline
$m^{\rm max}(q,\ell)$         &  362.1  &  297.7   \\
$\mctmax(q,q')$               &  396.9  &  396.9   \\
$\mctmax(\ell,\ell')$         &  151.9  &  ~69.7   \\
$\mctmax([q\ell][q'\ell'])$   &  727.2  &  668.5   \\
   \end{tabular}
}

A total of 800~K events were generated, corresponding to an integrated
luminosity of $\sim$12~fb$^{-1}$.  The generated events were passed
through the same parameterised detector simulation as for the top
sample described in Section~\ref{subsec4.1}. Events were selected with
the following requirements:
\begin{enumerate}
\item $N_{\rm jet}$ $\geq$ 2, with $p_T(j_1)$ $>$ 100 GeV and 
$p_T(j_2)$ $>$ 50 GeV.
\item $E_T^{miss}$ $>$ 100~GeV.
\item $N_{\rm lep}$ $=$ 2, where ${\rm lep} = e/\mu$(isolated) and
$p_T(l_2)$ $>$ 20 GeV. The two leptons were required to possess different
flavours.
\item Veto all events with jets with $p_T(j_1)$ $>$ 20~GeV labelled as a $b$-jet or $\tau$-jet.
\end{enumerate}
The veto on $b$ and $\tau$ labelled jets was applied to reduce the
SUSY background from events containing top quark or $\tau$ lepton
decays. 

The requirement of leptons with different flavour reduces the signal
by a factor two, but it is necessary, as the same-flavour signal is
dominated by SUSY background events in which the two leptons are
produced in the decay chain (\ref{decll}). This is demonstrated in
Figure~\ref{fig:OSSF} which shows the lepton-lepton invariant mass
distribution for opposite-sign same-flavour lepton pairs. The full
line is the inclusive distribution, and exhibits the characteristic
end-point structure from decay (\ref{decll}) \cite{Bachacou:2000zb};
the grey (red) area indicates lepton pairs from the decays
(\ref{decnu}) and (\ref{decl}). Figure~\ref{fig:OSSF} demonstrates
also that the characteristic lepton-lepton invariant mass end-point
from the \mbox{$\tchi^0_2\rightarrow\ell\tilde{\ell}$} decay chain is
indeed observable for this model.

\FIGURE[htb]{
\epsfig{file=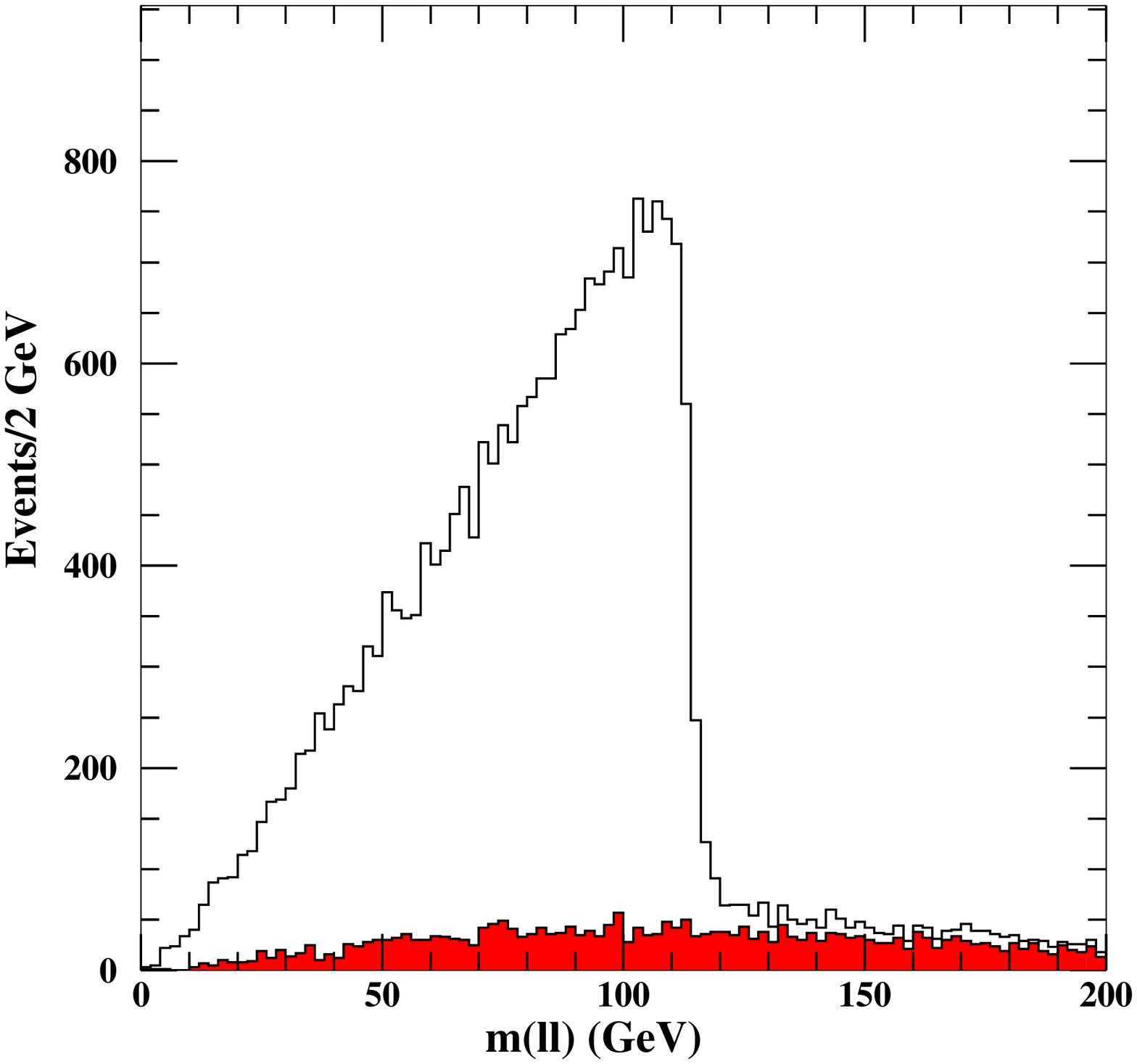,height=3.4in}
\caption{\label{fig:OSSF} Detector-level lepton-lepton invariant mass
for opposite-sign same-flavour lepton pairs after the selection cuts
except with the requirement of different flavours for the leptons.
The full line is the inclusive distribution while the grey (red) area
indicates the distribution for lepton pairs from the decays
(\ref{decnu}) and (\ref{decll}). } }

Following application of the selection cuts described above the only
significant remaining background was from $t\bar{t}$ production and only
this background is considered in the following. Approximately 9700
SUSY events passed cuts 1--4. Of these 7200 were indeed events in which
both the muon and the electron were produced directly in the decay of a
sparticle from the chains (\ref{decnu}) or (\ref{decl}). In the
remaining events at least one of the leptons was generated by the decay
of a tau lepton produced in one of the two legs of the event. The
number of $t\bar{t}$ background events was approximately 1400.

The parton-level distributions for the observables
$m(q^{(\prime)},\ell^{(\prime)})$,~~$\mct([q\ell],[q'\ell'])$,~~
$\mct(q,q')$ and~$\mct(\ell,\ell')$ are shown in Figure~\ref{fig:sqt}
for all events passing the selection cuts in which both legs in the
event contain the chain (\ref{decnu}) or the chain (\ref{decl}). The
contransverse mass observables have been corrected for transverse
boosts according to the procedure described in
Section~\ref{subsec2.2}. For the reasons discussed in
Section~\ref{subsec4.1} we shall not measure or exploit the
$\mct([q\ell],[q'\ell'])$ end-points in the following analysis, nor
the non-boost-corrected $\mct(b,b')$ and $\mct(\ell,\ell')$ versus
$p_b$ end-points. Nevertheless such constraints could be useful for
validating mass measurements obtained from the other observables.
\FIGURE[ht]{
\epsfig{file=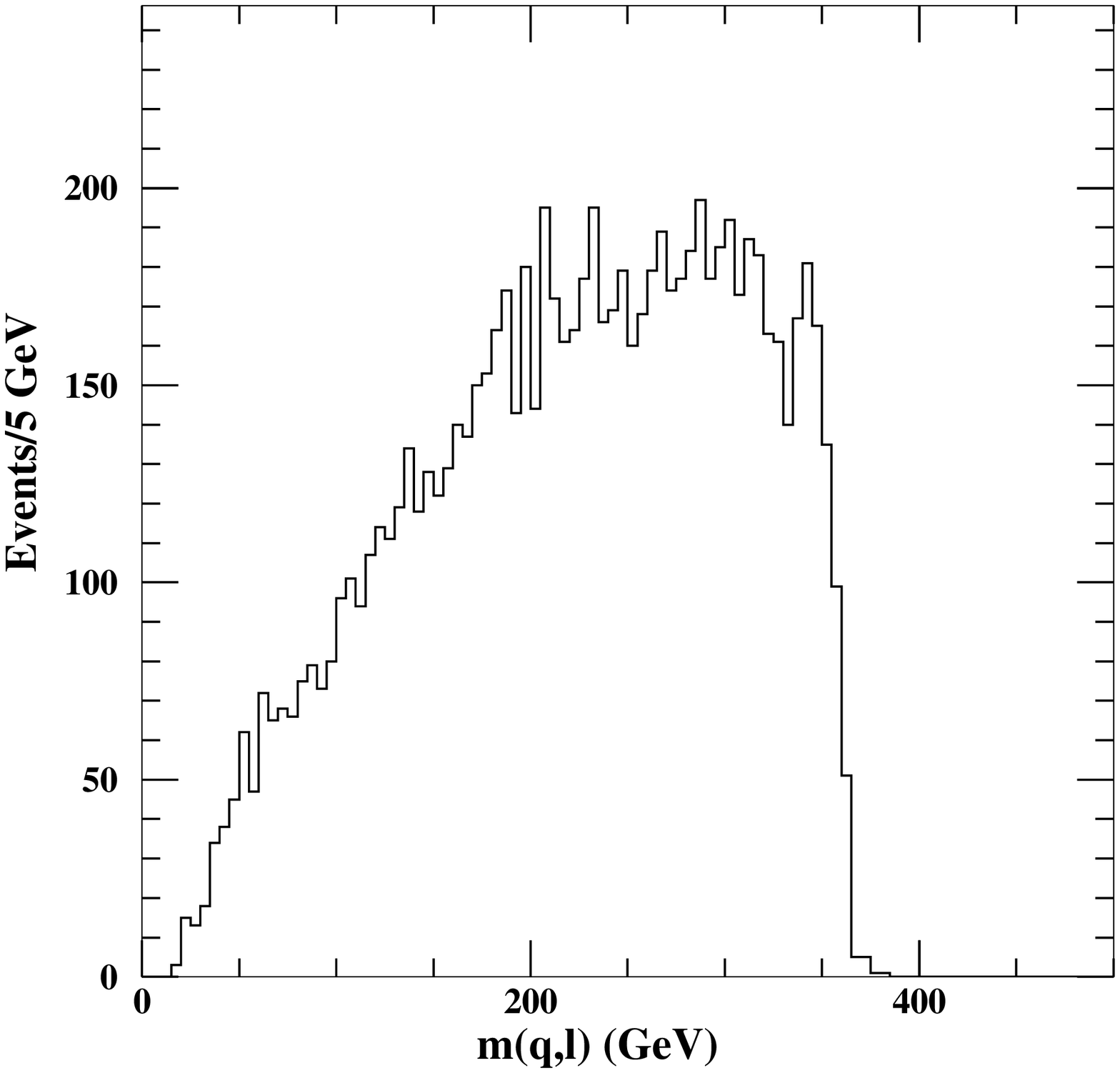,height=2.8in}
\epsfig{file=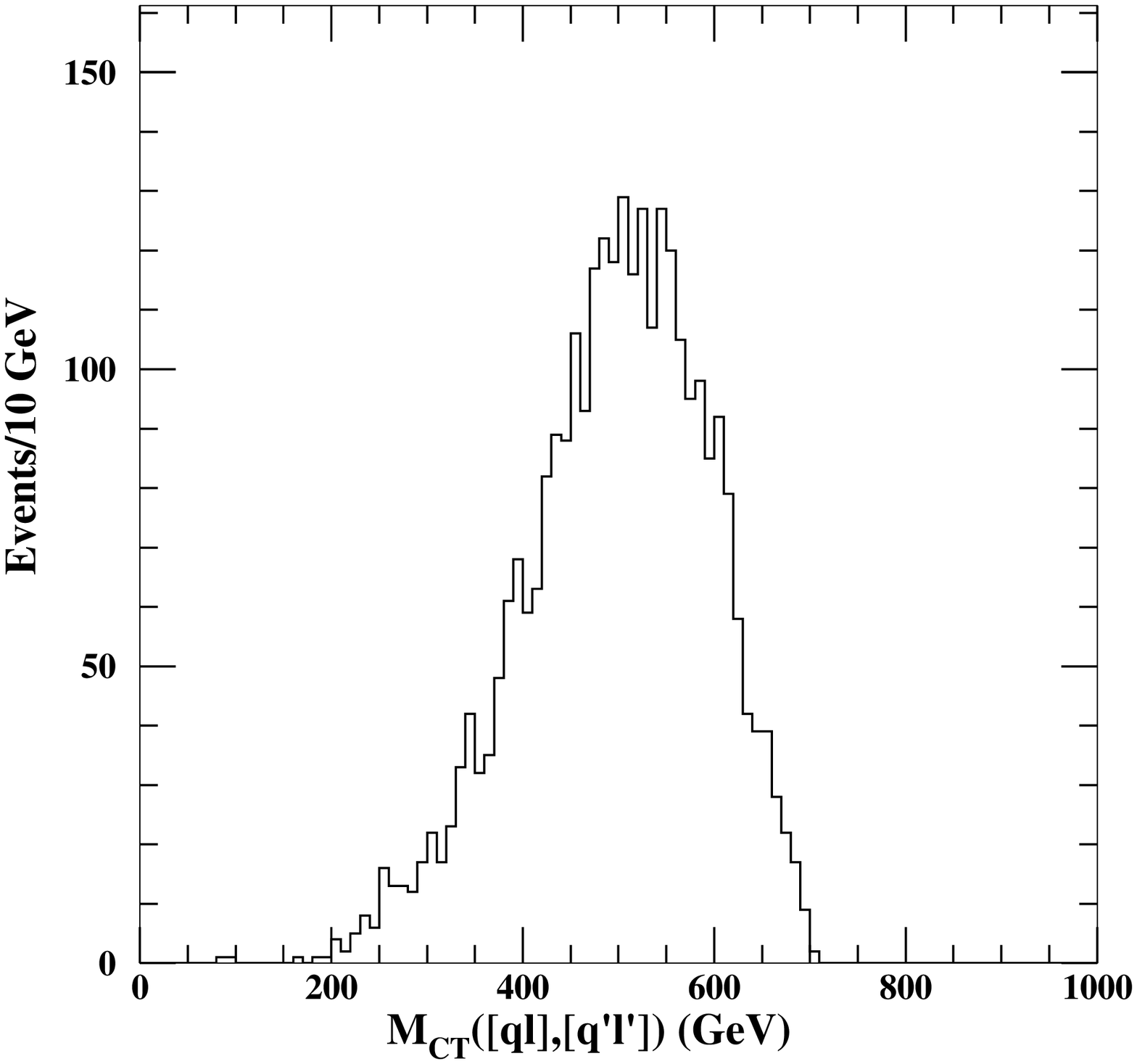,height=2.8in}
\epsfig{file=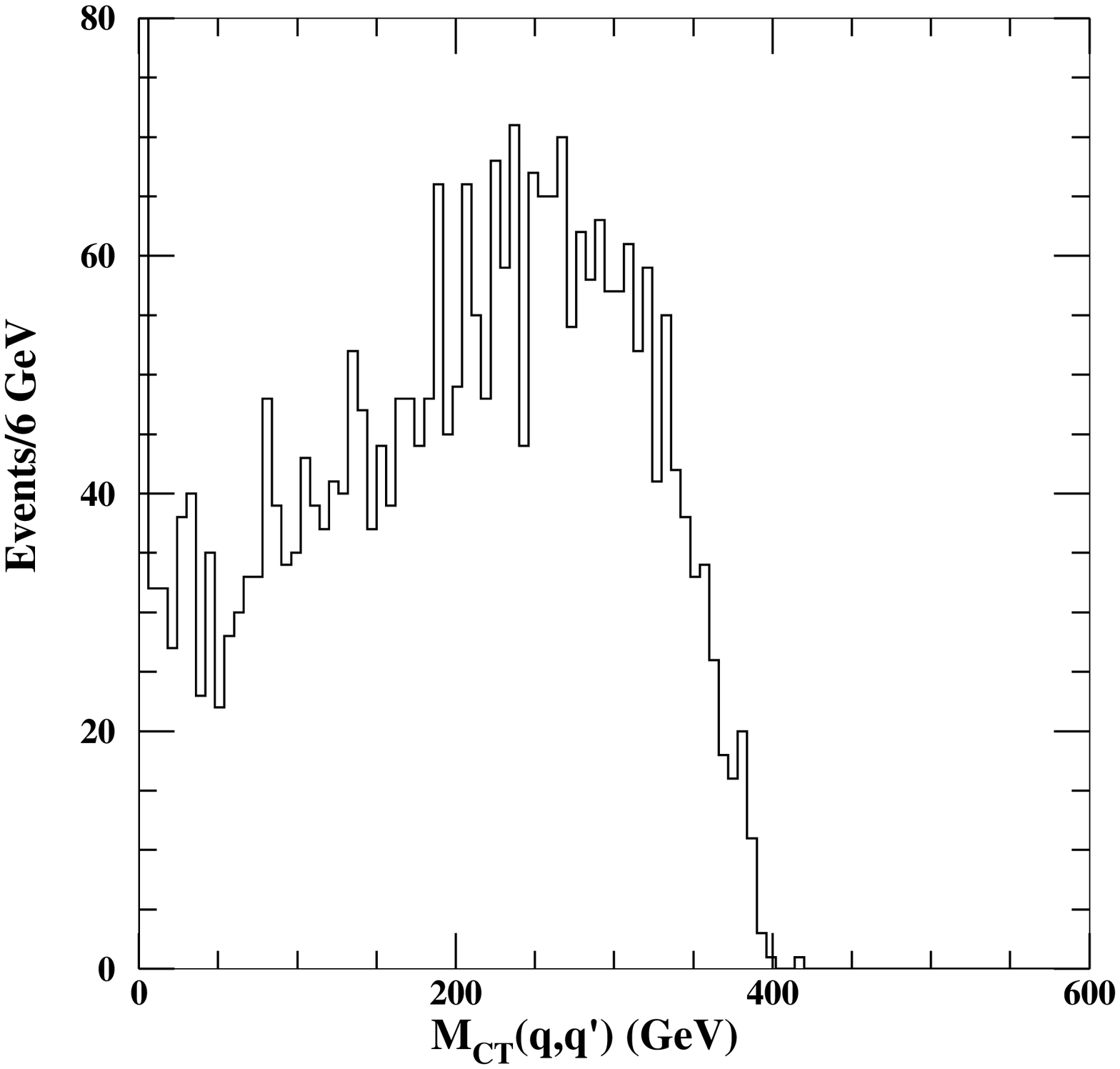,height=2.8in}
\epsfig{file=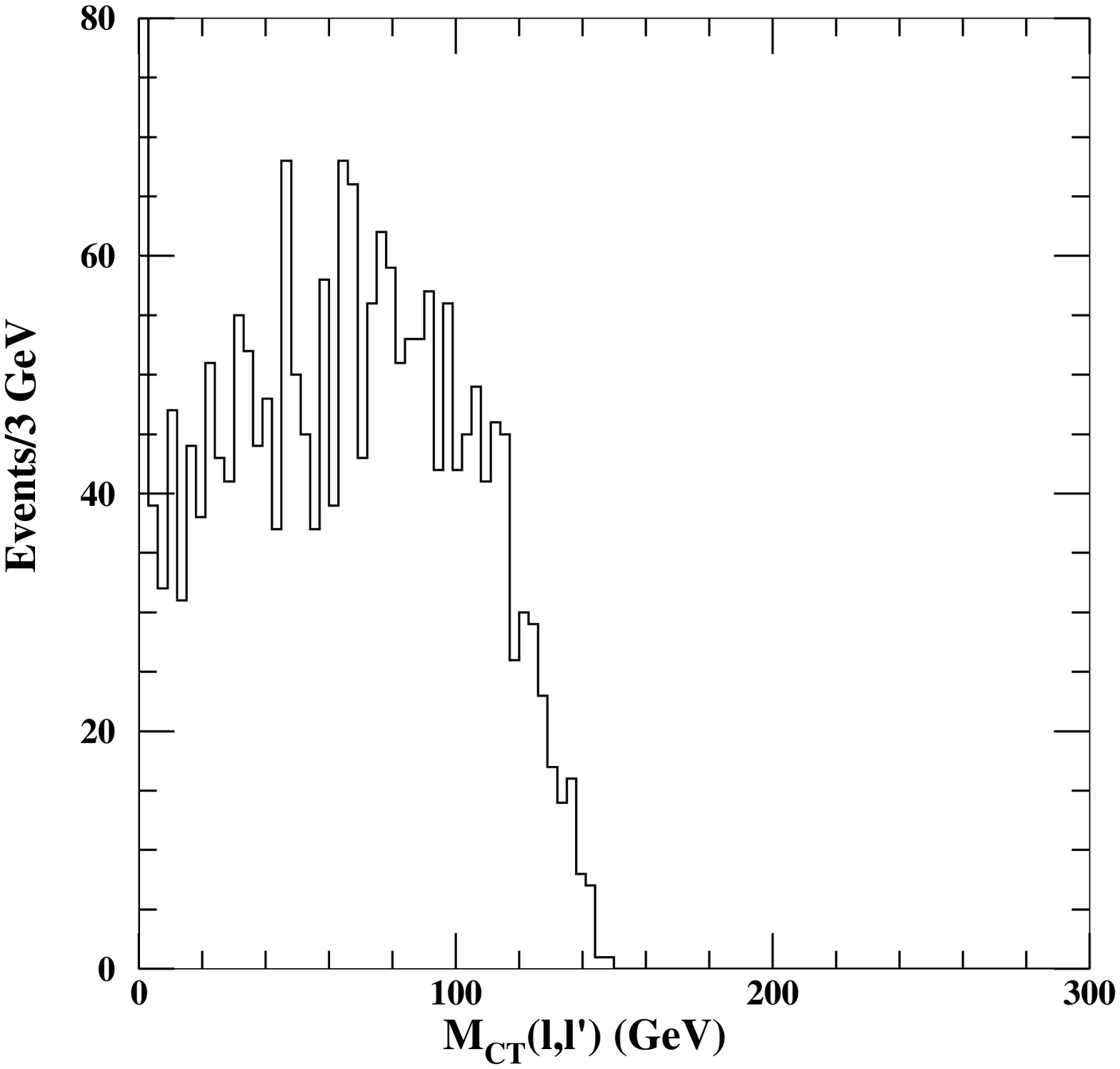,height=2.8in}
\caption{\label{fig:sqt} Parton-level distributions of $m(q^{(\prime)},\ell^{(\prime)})$ (top-left), $\mct([q\ell],[q'\ell'])$ (top-right), $\mct(q,q')$
(bottom-left) and $\mct(\ell,\ell')$ (bottom-right) for SUSY events
passing the selection cuts where both leptons are generated by decay
chain (\ref{decnu}). } }

The first step in the detector-level analysis is the calculation of
the invariant mass of each lepton with each of the two leading jets in
the event. The distribution of the minimum of these two masses for
each lepton is plotted in Figure~\ref{fig:mbls} and displays an
end-point at around 360~GeV, as expected from chain (\ref{decnu}). The
detector-level distributions of $\mct(q,q')$ and $\mct(\ell,\ell')$
are plotted in Figure~\ref{figlls} and also display end-points at the
positions expected for chain (\ref{decnu}). Only the distribution of
$\mct(\ell,\ell')$ values for events with $\axl<0$ is shown.  This is
due to the fact that the distribution at truth level for $\axl<0$ hits
the nominal end-point, whereas that for $\axl>0$ runs out of
statistics approximately 10~GeV below the nominal position (see
Figure~\ref{axplots}(left)), leading to a biased fitted end-point
position at detector level (see Figure~\ref{axplots}(right)). This
arises because only a very small boost correction (with $\eog=\ecm$)
can be applied in the $\axl>0$ case and hence the resulting corrected
$\mct$ value is more conservative than in the $\axl<0$ case. This
effect is most evident when considering$\mct(\ell,\ell')$ because of
the potentially large boosts generated by the recoiling $bb'$
system. The same effect, although numerically less evident, is present
also in the top quark analysis but is masked by the smearing of the
end-point due to the $W$ natural width. We choose here to use the
larger of the two fitted end-point positions, which must be nearer to
the true value.

\FIGURE[ht]{
\epsfig{file=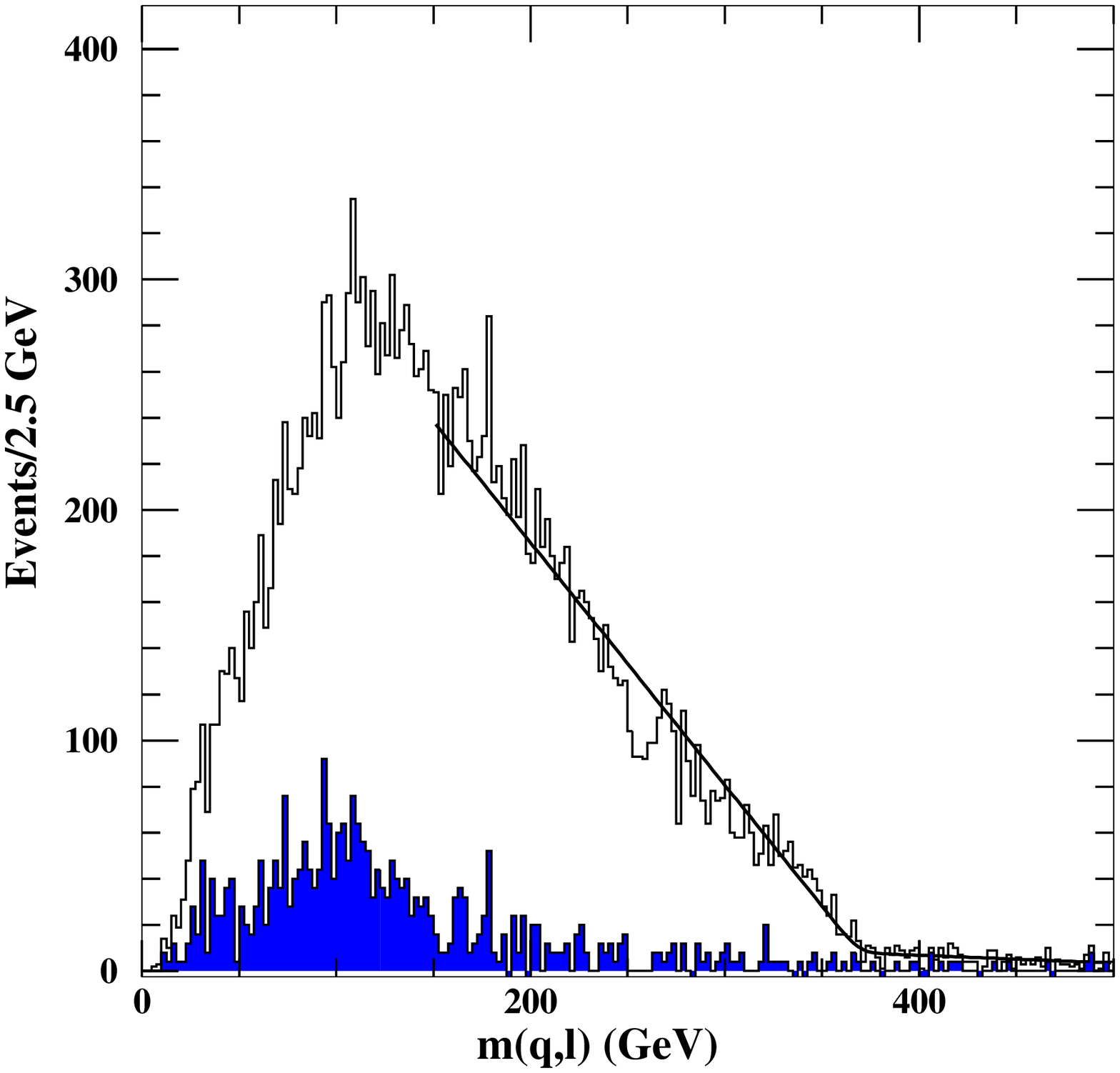,height=2.8in}
\epsfig{file=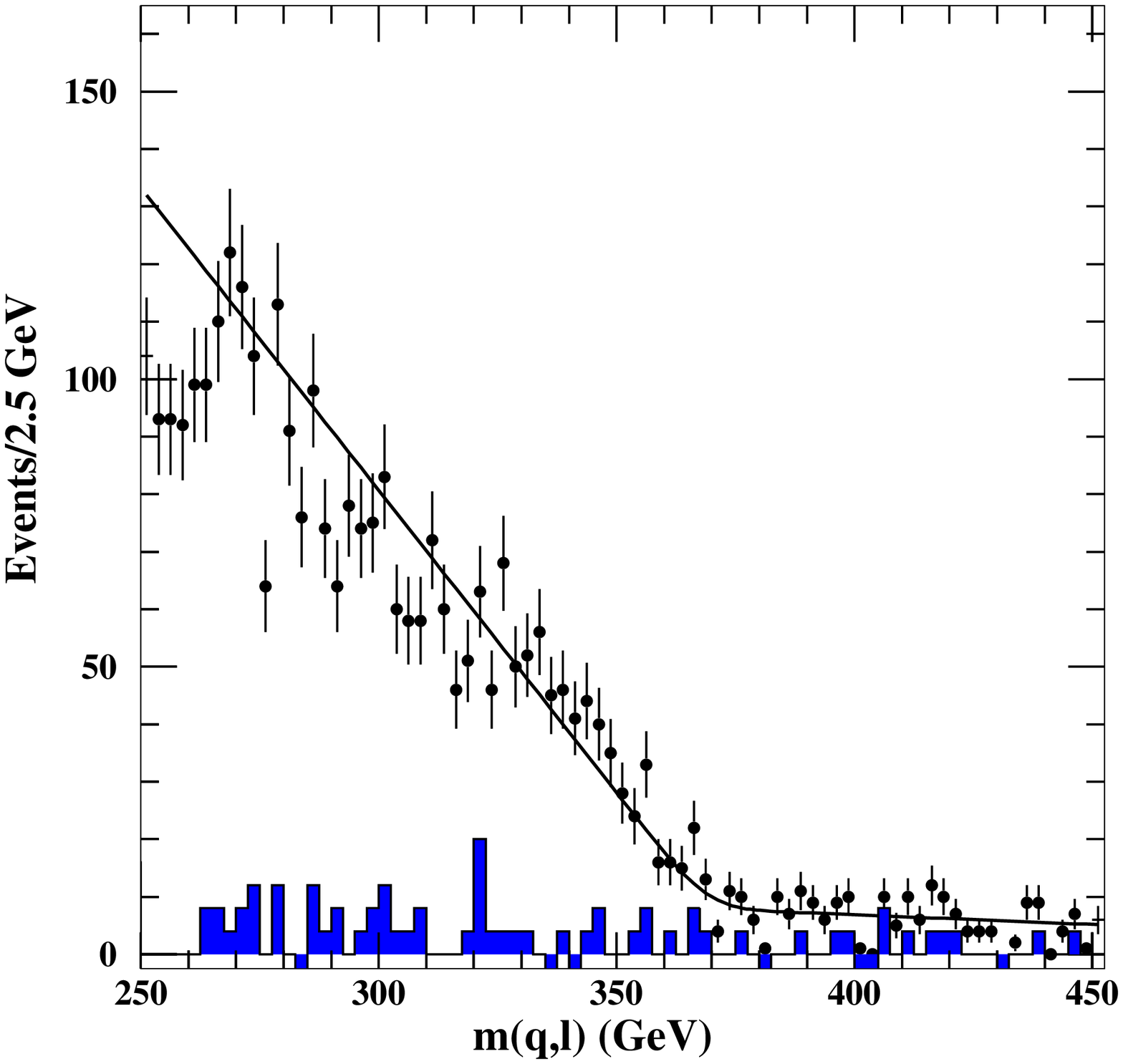,height=2.8in}
\caption{\label{fig:mbls} Detector-level distribution of
the minimum value of $m(q^{(\prime)},\ell^{(\prime)})$ for SUSY events
passing the selection cuts. The dark grey (blue) histogram indicates
the $t\bar{t}$ background. The complete distribution is shown on the
left while the region near the end-point is expanded on the right. The
fit to the end-point function given in Eqn.~(\ref{eq:fitfunctionllq})
is shown.}}

\FIGURE[ht]{
\epsfig{file=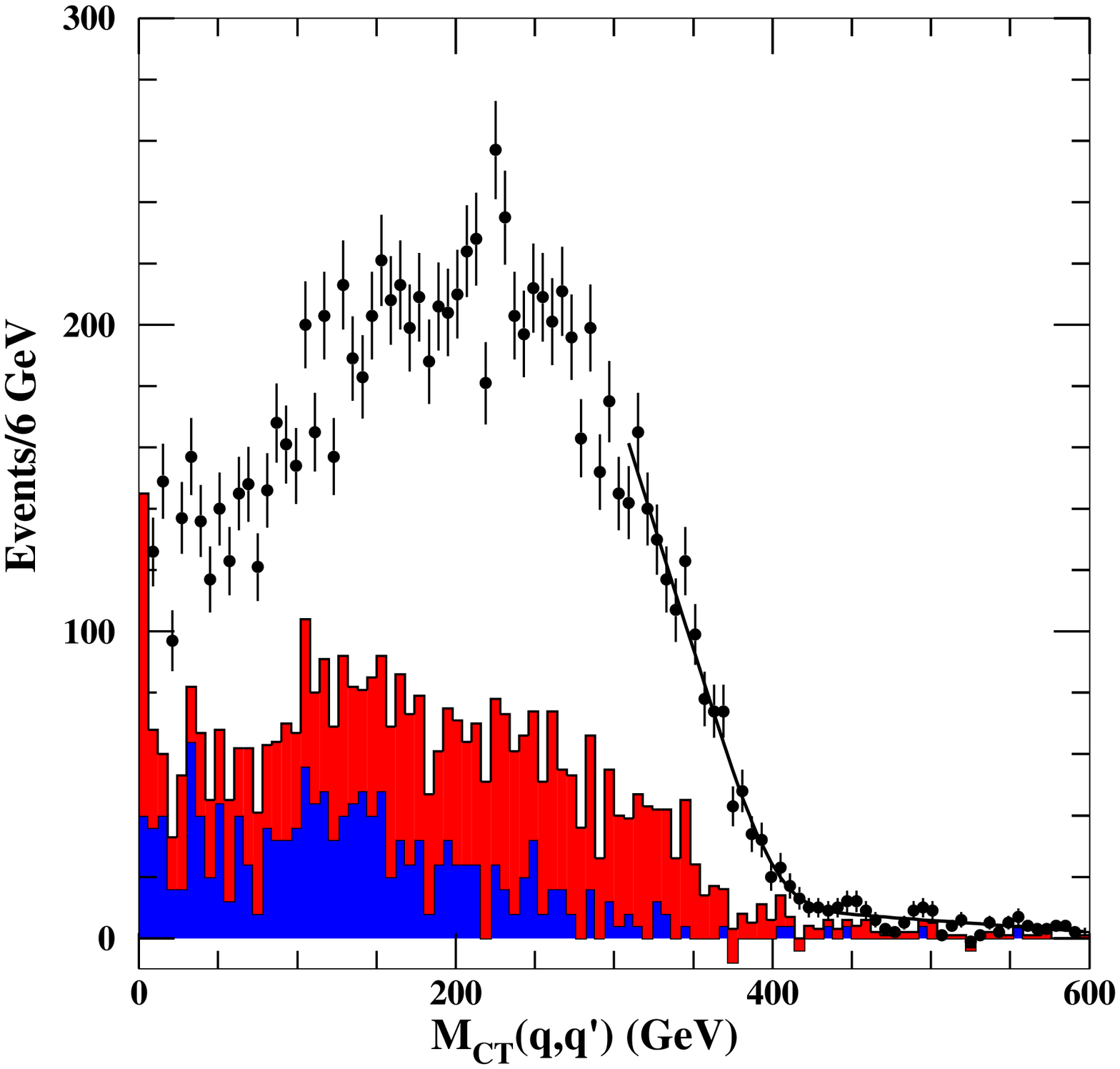,height=2.8in}
\epsfig{file=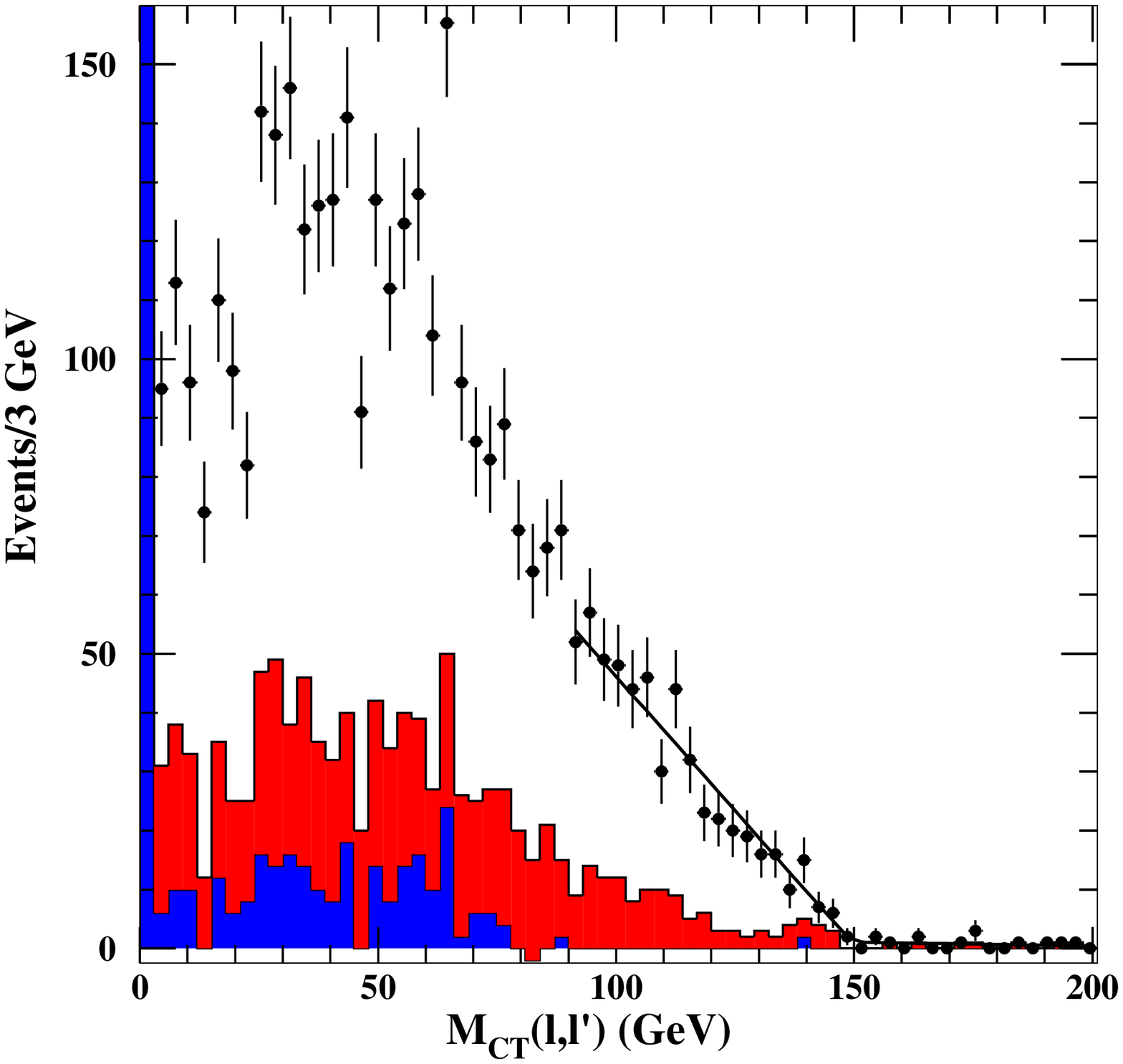,height=2.8in}
\caption{\label{figlls} Detector-level distributions of 
$\mct(q,q')$ (left) and $\mct(\ell,\ell')$ with $\axl<0$ (right) for
SUSY events passing the selection cuts. The light grey (red) histogram
is the distribution of events in which at least one of the two leptons
was not produced directly from the decay of a sparticle. The dark grey
(blue) area indicates the $t\bar{t}$ background. The fits to the
end-point function given in Eqn.~(\ref{eq:fitfunctionllq}) are shown. }
}

\FIGURE[ht]{
\epsfig{file=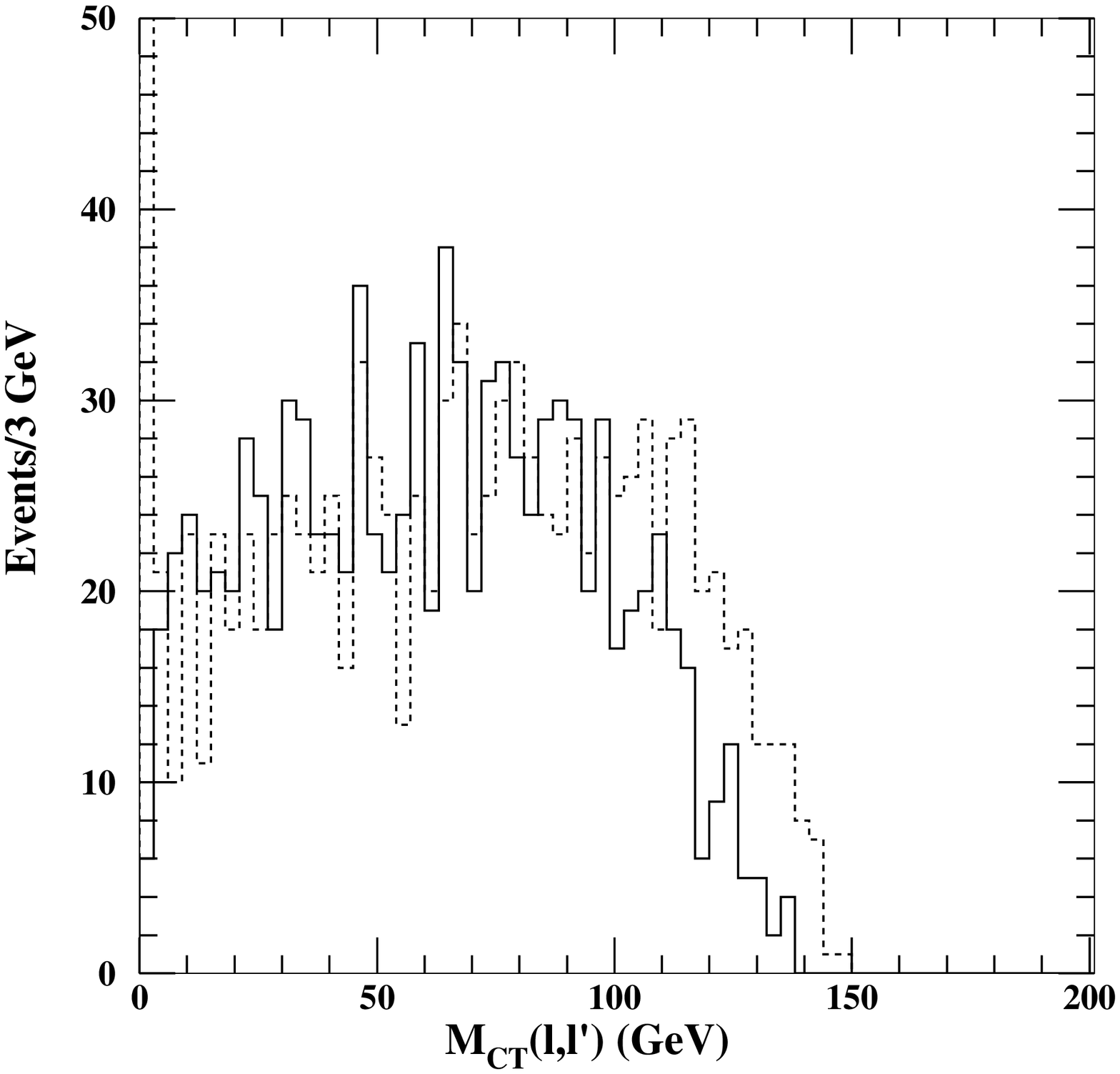,height=2.8in}
\epsfig{file=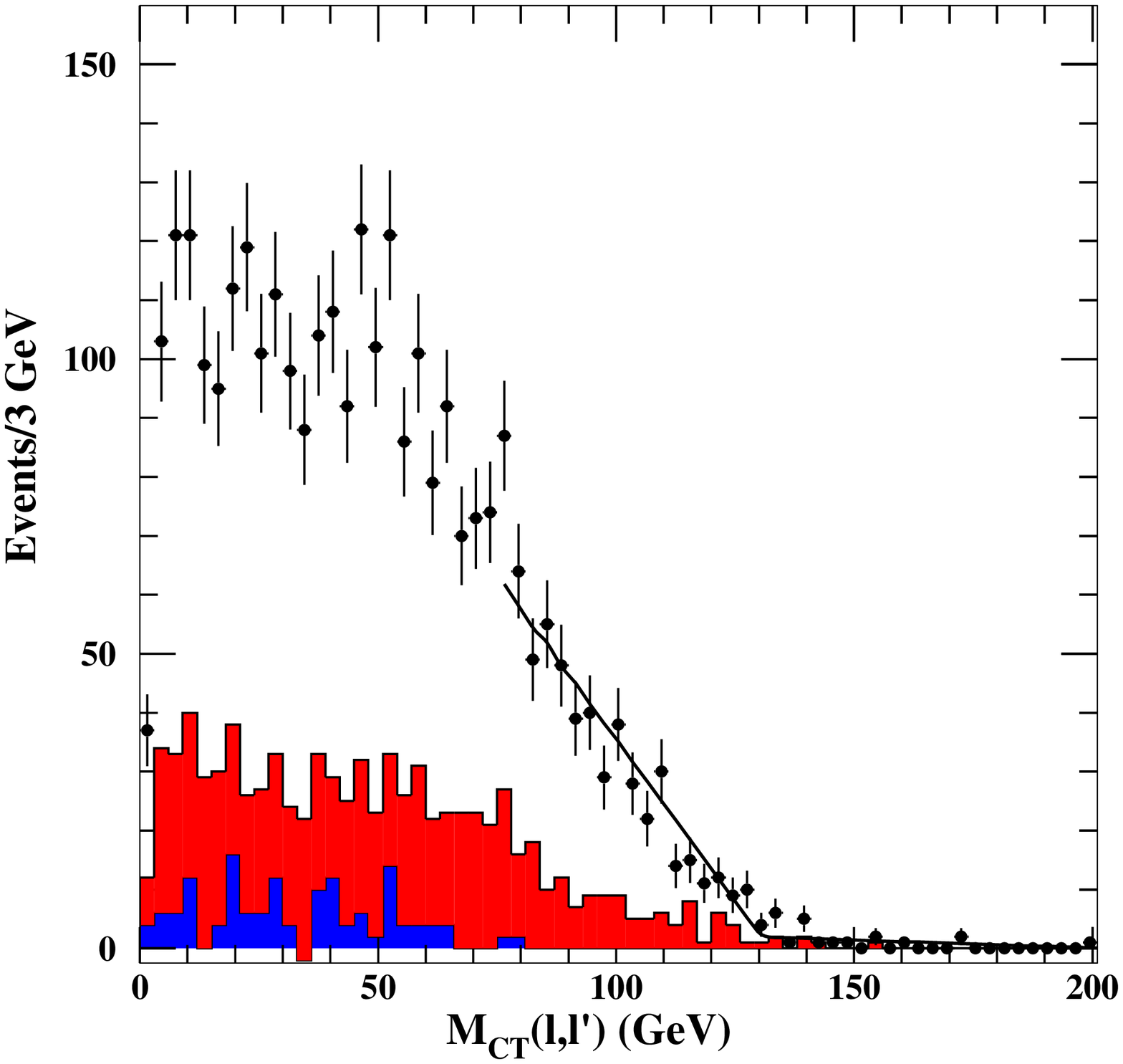,height=2.8in}
\caption{\label{axplots} Distributions of 
$\mct(\ell,\ell')$ at parton-level (left) and at detector-level
(right) for SUSY events passing the selection cuts. In the left-hand
figure the dashed histogram is the distribution of events with
$\axl<0$ while the full histogram is that of events with $\axl>0$. In
the right-hand figure all events possess $\axl>0$ and the shaded
histograms and fitted curve are as for Figure~\ref{figlls}. The fitted
end-point position is 133.6 GeV.} }

In order to explore the approximate potential precision of mass
measurements obtained with this technique, we fit the end-points of
the distributions with the smeared linear function given in
Eqn.~(\ref{eq:fitfunctionllq}). The caveats associated with this
technique discussed in Section~\ref{subsec4.1} are also relevant
here. In the case of the $\mct(\ell,\ell')$ distribution we only fit
the distribution of events with $\axl<0$, as discussed above. The
fitted distributions are shown in Figures~\ref{fig:mbls}
and~\ref{figlls}.
\TABLE[ht]{
\caption{End-point positions in GeV. The first uncertainty is
statistical, while the second and third are respectively the
uncorrelated systematic and correlated energy scale uncertainties. The
expected end-point positions from Eqns.~(\ref{eqn20a}),~(\ref{eqn20b})
and~(\ref{eqn20c}) are listed in the column labelled `Truth'. The
assumed integrated luminosity is 12~fb$^{-1}$}
\bigskip
\label{tabfits}
   \begin{tabular}{l|c|c}
   End-point &  Truth (GeV) & Measured  (GeV) \\
   \hline
$m^{\rm max}(q,\ell)$           & 362.1  &  $ 369.2 \pm 2 \pm 5 \pm 1.5$ \\
$\mctmax(q,q')$                 & 396.9  &  $ 401.7 \pm 4.8 \pm 5 \pm 4$ \\ 
$\mctmax(\ell,\ell')$ $(A_x<0)$ & 151.9  &  $149.3 \pm 1.5 \pm 3 \pm 0.8$ \\
   \end{tabular}
}
The results of the end-point fits are listed in Table~\ref{tabfits},
where the first uncertainty is the statistical uncertainty from the
{\tt MINUIT} \cite{James:1975dr} fitting program for the chosen fit
interval, the second is the systematic uncertainty obtained by varying
the fit interval and the third uncertainty is the correlated
systematic uncertainty derived from assumed energy scale uncertainties
of 1\% for jets and 0.1\% for leptons \cite{atltdr}. As for the top
study the quoted uncertainties should be considered approximate and
could be improved with the use of better end-point fitting functions,
for instance templates derived from Monte Carlo simulation studies.

If we use the measured end-point positions to calculate the masses of
the sparticles using Eqns.~(\ref{eq:ma}),~(\ref{eqn20b})
and~(\ref{eqn20c}) we obtain an uncertainty of 20~GeV on the absolute
squark mass, an uncertainty of 6~GeV on the difference between the
squark mass and the masses of the other sparticles, and an uncertainty
of 3~GeV on the chargino-sneutrino mass difference. We have thus shown
with a toy SUSY model that it is possible to achieve a stand-alone
measurement of sparticle masses using the contransverse mass technique
applied to events containing two symmetric sequential two-step
two-body decay chains.

%% file: conclusions.tex
\section{Conclusions}\label{sec5}
In this paper we have extended the contransverse mass technique for
measuring the masses of pair-produced semi-invisibly decaying heavy
particles so that it can be applied to events with non-negligible
boosts of the CoM frame of the heavy states in the laboratory
transverse plane. We have demonstrated the modified technique with
case studies measuring the masses of the top quark, $W$ and neutrino
in fully leptonic $t\bar{t}$ events, and the masses of sparticles in
SUSY events with a similar final state. The case studies presented
here are in many respects more detailed than previous studies of
alternative strategies and illustrate well the potential utility of
the contransverse mass technique.

%% file: appendix.tex
\renewcommand{\theequation}{A\arabic{equation}}
\setcounter{equation}{0}
\section{The connection between $M_{T2}(\chi)$ and $\mct$}

Eqn.~(\ref{eqn12}) can be used to study the links between
$M_{T2}(\chi)$ \cite{Lester:1999tx} and $\mct$, as we shall now
illustrate. In the process we shall obtain an approximate analytical
expression for a boost-corrected version of $M_{T2}(\chi)$. The link
between the two quantities in the absence of co-linear transverse
boosts and for massless visible states was first discussed in
Ref.~\cite{Serna:2008zk}.

First, observe that if we assume a value for $m(\alpha)$ and know
$\mctmax$ then we can solve Eqn.~(\ref{eqn12}) for $m(\delta)$. For
any given event we do not know $\mctmax$ however, but rather
$\mct$. Let us therefore substitute $\mct$ for $\mctmax$ to obtain the
following solution
\begin{equation}
\label{eqn13a}
m_{\rm soln}(\delta) = \Big(\chi^2 + A_T +
\sqrt{\Big[1+\frac{4\chi^2}{2A_T-m^2(v_1) -
m^2(v_2)}\Big]\Big[A_T^2-m^2(v_1)m^2(v_2)\Big]}\Big)^{1/2},
\end{equation}
where $A_T \equiv [\mct^2-m^2(v_1)-m^2(v_2)]/2$ and $\chi$ is the
assumed value of $m(\alpha)$. This is identical to the analytical
expression for the `balanced' solution of $M_{T2}(\chi)$ in the
absence of co-linear transverse boosts, which was identified in
Refs.~\cite{Lester:2007fq,Cho:2007dh}. In general $m_{\rm
soln}(\delta)$ need not be bounded by $m(\delta)$, because $A_T$ and
hence $\mct$ appears in the denominator inside the square-root causing
$m_{\rm soln}(\delta)$ not to be a monotonically increasing function
of $\mct$. Therefore a value of $\mct<\mctmax$ need not
generate a value of $m_{\rm soln}(\delta)$ which is less than the
value obtained with $\mct=\mctmax$. Note that if $\chi=0$ then $m_{\rm
soln}(\delta)$ is nevertheless bounded by $m(\delta)$ because in this
special case $m_{\rm soln}(\delta)$ {\it is} a monotonically increasing
function of $\mct$.
 
In general $m_{\rm soln}(\delta)$ is not invariant under co-linear
transverse boosts, because it depends on $\mct$ which is also not
invariant. One might consider therefore whether it is possible to
correct $m_{\rm soln}(\delta)$ for such boosts in a similar manner to
that used to correct $\mct$ in Section~\ref{subsec2.2}. The $\mct$
boost-correction procedure minimises $\mct$ with respect to the
possible boosts, however this does not necessarily minimise $m_{\rm
soln}(\delta)$ because it is not in general a monotonically increasing
function of $\mct$. Boost-correction can be performed however, as we
know not only the minimum possible value of $\mct$ in the
$\delta_1\delta_2$ CoM frame but also the maximum possible value,
which is given by the maximum of the $\mct$ values obtained by
boosting $v_1$ and $v_2$ with respectively $\beta=p_b/E_{\rm cm}$ and
$\beta=p_b/\ehat$ (see Fig.~\ref{fig1}). We therefore know the range
of CoM frame $\mct$ values which could have occurred in the event.

To proceed we find the turning-points of Eqn.~(\ref{eqn13a}), which
lie at the following four values of $\mct^2$:
\begin{equation}
\mct^2 = \frac{\chi[3m^2(v_i)+m^2(v_j)]\pm2m(v_i)[m^2(v_i)+m^2(v_j)]}{\chi\pm m(v_i)},
\end{equation}
where $i,j=1,2$ $(i\neq j)$. We then take the minimum of all the $m_{\rm
soln}(\delta)$ values obtained at turning-points where $\mct$ lies
within the allowed range identified above. We finally take the minimum
of this $m_{\rm soln}(\delta)$ value and those obtained with $\mct$
set to its extrema. The result is a boost-corrected value of $m_{\rm
soln}(\delta)$.

Now although $m_{\rm soln}(\delta)$ is not in general bounded by
$m(\delta)$ the full analytical expression for $M_{T2}(\chi)$ in the
absence of co-linear transverse boosts \cite{Lester:2007fq,Cho:2007dh}
is so bounded. This is because in those cases where $m_{\rm
soln}(\delta)>m(\delta)$ $M_{T2}(\chi)$ takes on an alternative value
corresponding to an `unbalanced' solution (see e.g. Eqn.~(54) in
Ref.~\cite{Lester:2007fq}). The quantities upon which the decision to
switch to such an alternative value rests are not themselves
boost-invariant. Development of an appropriate boost-correction
procedure for these quantities consistent with the parallel
boost-correction of $m_{\rm soln}(\delta)$ requires more work. In the
absence of such a correction procedure we can choose to neglect the
boost when making this decision, to obtain an approximate analytical
form for a boost-corrected version of $M_{T2}(\chi)$. 

Note that the quantity calculated here is not the same quantity as the
conventional $M_{T2}(\chi)$ used elsewhere, which is boost-independent
but currently does not possess a general analytical form. The
boost-correction procedure, even if exact (i.e in the absence of the
approximation mentioned in the previous paragraph), leads to a
quantity which is not in general equal to the conventional
$M_{T2}(\chi)$, although equality is obtained in the absence of
co-linear transverse boosts.

The procedure described above for calculating this `boost-corrected
$M_{T2}(\chi)$' quantity is implemented in the boost-correction package
described in Section~\ref{subsec2.2}, available at {\tt
http://projects.hepforge.org/mctlib}.

\renewcommand{\theequation}{B\arabic{equation}}
\setcounter{equation}{0}
\section{Invisible pseudo-particle masses for chargino decay chains}
If a heavy sparticle decays via a chain which produces multiple
invisible final state particles then the values of $m^{\rm max}(P,Q)$
and $\mctmax([PQ],[P'Q'])$ can be calculated by constructing an
aggregate `pseudo-particle' $\alpha$ from the invisible particles. The
minimum value of $m(\alpha)$, $m_{\rm min}(\alpha)$, can then be used
in end-point formulae, as described in Section~\ref{subsec4.2}.

If a chargino decays through $\chipm \ra \chioi W^{\pm} \ra \chioi
\nu \ell^{\pm}$ then $\alpha \equiv [\chioi\nu]$ and $m_{\rm
min}(\alpha)$ is given by:
\begin{equation}
m_{\rm min}^2(\alpha)=\mchioisqr+\mw\big[E(\chioi)-p\big]\sqrt{\frac{E(W)-p}{E(W)+p}},
\end{equation}
where
\begin{eqnarray}
p &\equiv& \frac{\sqrt{[\mchipmsqr-\mchioisqr-\mwsqr]^2-4\mchioisqr\mwsqr}}{2\mchipm},\\
E(W) &=& \frac{\mchipmsqr-\mchioisqr+\mwsqr}{2\mchipm}, \\
E(\chioi) &=& \frac{\mchipmsqr+\mchioisqr-\mwsqr}{2\mchipm},
\end{eqnarray}
and we have assumed $\mnu=0$. If however the chargino decays
through $\chipm \ra \nu \slep \ra \nu \ell^{\pm}\chioi$, as in decay
chain (\ref{decl}), then:
\begin{equation}
m_{\rm min}(\alpha) = \frac{\mchipm\mchioi}{\msl},
\end{equation}
while if the chargino decays through $\chipm \ra \ell^{\pm} \snu \ra
\ell^{\pm}\nu \chioi$, as in decay chain (\ref{decnu}), then $\alpha
\equiv \tilde{\nu}$ and $m_{\rm min}(\alpha)$ is fixed to
$m(\alpha)$ given by:
\begin{equation}
m(\alpha) = \msnu.
\end{equation}
Finally, if the chargino decays through the three-body decay $\chipm
\ra \chioi \ell^{\pm}\nu$ then $\alpha \equiv [\chioi\nu]$ and $m_{\rm
min}(\alpha)$ is given by:
\begin{equation}
m_{\rm min}(\alpha) = \mchioi.
\end{equation}